\documentstyle[12pt]{article}                                         

\newcommand{\ba}{\begin{eqnarray}}   
\newcommand{\ea}{\end{eqnarray}}     
\newcommand{\be}{\begin{equation}}   
\newcommand{\ee}{\end{equation}}     

\newcommand{\RMA}{{\cal R}}     
\newcommand{\Acr}{{\cal A}}     
\newcommand{\ALG}{{\cal A}}     
\newcommand{\HH}{{\cal H}}      
\newcommand{\CC}{{\rm C}}       
\newcommand{\RR}{{\rm R}}       
\newcommand{\Id}{{\rm I}}       
\newcommand{\PP}{{\rm P}}       
\newcommand{\hh}{h}             
\newcommand{\HAM}{H}            
\newcommand{\ham}{h}            
\newcommand{\VV}{V}             
\newcommand{\pp}{p}             
\newcommand{\PMO}{P}            
\newcommand{\al}{\alpha}
\newcommand{\bet}{\beta}
\newcommand{\ga}{\gamma}
\newcommand{\Ga}{\Gamma}
\newcommand{\dl}{\delta}
\newcommand{\ep}{\epsilon}
\newcommand{\Dl}{\Delta}
\newcommand{\si}{\sigma}
\newcommand{\ka}{\kappa}
\newcommand{\lam}{\lambda}
\newcommand{\om}{\omega}
\newcommand{\Om}{\Omega}
\newcommand{\veps}{\varepsilon}
\newcommand{\vph}{\varphi}

\newcommand{\arctg}{{\rm arctg}}
\newcommand{\tr}{{\rm tr}}      %
\newcommand{\odd}{{\rm odd}}    %
\newcommand{\even}{{\rm even}}  %
\newcommand{\Imm}{{\rm Im}}     %
\newcommand{\Spher}{{\rm S}}    

\newcommand{\sltwo}{sl(2)}                
\newcommand{\slqtwo}{sl_{q}(2)}           
\newcommand{\slaff}{\widehat{sl_{q}}(2)}  
\newcommand{\sltwor}{sl(2,\RR)}           
\newcommand{\ofour}{o(4)}                 %
\newcommand{\Ethree}{E(3)}                %

\newcommand{\leqslant}{\leq}    
\newcommand{\geqslant}{\geq}    

\begin{document}
\vskip2.cm
\centerline{\bf
        How Algebraic Bethe Ansatz works for
            integrable model}

\vskip1.cm
\centerline{\bf
    L.~D.~Faddeev}

\vskip1.cm
\centerline{\it
St.Petersburg Branch of Steklov Mathematical Institute}          
\centerline{\it
Fontanka 27, St.Petersburg 191011}                                
\centerline{\it
Russia}                                                             
\vskip0.5cm
\centerline{\it
Research Institute for Theoretical Physics}
\centerline{\it
University of Helsinki}
\centerline{\it
Siltavuorenpenger 20C}
\centerline{\it
SF 00014 Helsinki}
\centerline{\it
Finland}

\vskip10.cm
\centerline{\bf 1995}                                                          

\newpage
\section{ Introduction }
\label{sone}

                In my Les--Houches lectures of 1982 I described
                the inverse scattering method of solving the
                integrable field--theoretical models in
        $ 1+1 $
                dimensional space--time. Both classical case,
                stemming from the famous paper by Gardner,
                Green, Kruskal and Miura of 1967 on KdV
                equation, and its quantum counterpart,
                developed mostly by Leningrad group around
                1978--79, were discussed. In particular, the
                algebraic way of deriving the Bethe--Ansatz
                equations was presented, but its use was not
                illustrated in any detail. I just stated in
                the end of the lectures, that ``The hard work
                just begins''.

                In this course I shall exactly describe this
                work. During last 10 years I lectured several
                times on this subject, but this particular
                lecture course is the longest and more
                detailed.

                In the announcement of the School my course is
                called ``Hidden symmetries in integrable
                models''.
                The term ``symmetry'' in modern
                literature on mathematical physics is supplied
                with several adjectives, such as hidden,
                dynamical, broken, deformed etc., but there is
                no exact definition for all this. Thus I
                decided to change my title in the proceedings
                to reflect more adequately the actual content
                of the course.

                The term ``integrable models'' in the title
                refers to particular family of quantum
                field--theoretical models in
        $ 1+1 $
                dimensional space--time, which are solvable by
                means of the quantum variant of inverse
                scattering method. The adjective ``integrable''
                stems from a paper of Zakharov and me of 1971
                where the KdV equation was shown to allow the
                interpretation as an integrable (though
                infinite--dimensional) hamiltonian system.

                The most famous example of the integrable
                model is the Sine--Gordon equation
\be
        \Box \vph + m^2 \sin \vph = 0
\ee
		for the scalar field
        $ \vph(x,t) $,
		which is relativistic and nonlinear.

                As is well known, the quantum field--theoretic
                models with interaction are plagued by
                infinities and some regularization is
                necessary. The discretizing of the space, or
                going to the lattice, is one way for such a
                regularization. It reduces the field model in
                the finite volume to a system with finite
                number of degrees of freedom.

                In the beginning of 80--ties, due mainly to the
                work of Izergin and Korepin it was realized,
                that the integrable models allow the lattice
                counterparts, which are also integrable and can
                be interpreted as the quantum spin models of
                magnetic chains. This unexpected but very
                welcome connection showed the universality of
                the spin chains in the domain of integrable
                models. In my course I shall begin with and
                speak in detail on the spin chains. The
                field--theoretical models will appear as their
                particular continuous space limits.

		One can ask, what is good in
	$ 1+1 $
                models, when our space--time is
        $ 3+1 $--dimensional.
                There are several particular
                answers to this question.

                1. The toy models in
	$ 1+1 $
                dimension can teach us about the realistic
                field--theoretical models in a nonperturbative
                way. Indeed such phenomena as renormalization,
                asymptotic freedom, dimensional transmutation
                (i.e. the appearance of mass via the
                regularization parameters) hold in integrable
                models and can be described exactly.

                2. There are numerous physical applications of
                the
        $ 1+1 $
                dimensional models in the condensed matter
                physics.

                3. The formalism of integrable models showed
                several times to be useful in the modern
                string theory, in which the world sheet is
                2--dimensional anyhow. In particular the
                conformal field theory models are special
                massless limits of integrable
                models.

                4. The theory of integrable models teaches us
                about new phenomena, which were not appreciated
                in the previous developments of Quantum Field
                Theory, especially in connection with the mass
                spectrum.

                5. I cannot help mentioning that working with
                the integrable models is a delightful pastime.
		They proved also to be very successful tool for
                the educational purposes.


                These reasons were sufficient for me to
                continue to work in this domain for last 25
                years (including 10 years of classical
                solitonic systems) and teach quite a few of
                followers, often referred to as
                Leningrad--St.Petersburg school.

                I am very grateful to the organizers of the
                school Professors A.Connes and K.Gawedsky for
                inviting me to lecture. First, it is always
                nice to be in Les--Houches and it is already my
                third lecture course here. Second, the texts
                of two previous lectures were transformed into
                monographs later. I hope that this course will
                eventually lead to one more book dedicated to
                the quantum theory of solitons.

\section{ General outline of the course. }
\label{stwo}
                My usual methodological trick in teaching is
                not to begin in full generality but rather to
                choose a representative example and explain on
                it all technical features in such a way, that
                generalization become reasonably evident. Thus
                I begin the course with the concrete example
                of the magnetic model -- the spin
        $ 1/2 $
		XXX chain.

                All the ingredients of Algebraic Bethe Ansatz,
                which is another name for the Quantum Inverse
                Scattering method, namely Lax operator,
                derivation of Bethe--Ansatz equations,
                thermodynamic limit, will be presented in
                full detail on this example. Then the
                generalization to XXX model of spin
        $ s > 1/2 $
                and XXZ model will follow. Here only features,
                which distinguish this model from the basic
                one, will be described. Finally the continuous
                field--theoretical models such as Nonlinear
                Schroedinger Equation,
        $ \Spher^2 $
		nonlinear
	$ \sigma $--model,
                principal chiral field model for
	$ SU(2) $
                and Sine--Gordon model will be included as
                particular limits of spin chains.

                We will see, that in the description of
                dynamics the finite time shift
\be
	U = e^{-iH\Delta}
\ee
                will appear naturally. This will make our
                discretization scheme more consistent, time
                and space being discrete simultaneously.

                Now I shall present some kinematics of the
                models we shall consider.

                As ``space'' we shall consider a discrete
                circle, namely the ordered set of points,
                labeled by integers
        $ n $
		with the identification
	$ n \equiv n+N $,
		where
	$ N $
                is a fixed positive integer.
                As a ``fundamental domain'' we shall take
        $ n = 1, \dots , N $.
		The integer
	$ N $
                plays the role of the volume of the space; the
                identification reflects the periodic boundary
                condition.

                Formal continuous limit will be described by
                introducing the lattice spacing
        $ \Delta $
		and coordinate
	$ x = n \Delta $
		which becomes continuous in the limit
	$ \Delta \rightarrow 0 $, $ N \rightarrow \infty$.
		In particular the following rule will be used
\be
        \frac{\delta_{mn}}{\Delta} = \delta (x-y) \quad ,
\ee
		so that the Kroneker symbol
	$ \delta_{mn} $
		is of order
	$ \Delta $.

		The quantum algebra of observables
        $ \ALG $
		is generated by dynamical variables
	$ X^{\alpha}_n $,
		attached to each lattice site
	$ n $.
		Index
	$ \alpha $
                assumes some finite number of values.
                The algebra
        $ \ALG $
                is defined by fixing the set of commutation
                relations between
	$ X^{\alpha}_n $.
		These relations are called ultralocal, when
	$ X^{\alpha}_m $ and $ X^{\beta}_n $
		commute for
	$ n \neq m $.
		A more relaxed condition of locality is that
	$ X^{\alpha}_m $ and $ X^{\beta}_n $
		do not commute only for
	$ | n - m | $
		small, in particular for
	$ n = m - 1 $ and $ n = m + 1 $ .

                Let us present examples,
                beginning with ultralocal case.
\begin{enumerate}

  \item         Canonical variables
        $ \vph^\alpha_n , \pi^\alpha_n , \alpha = 1,\dots,l $
		with the relations
\ba
      & [ \vph^{\alpha}_n , \vph^{\beta}_m ]   =   0 \; ,
		\;\;
        [ \pi^{\alpha}_n , \pi^{\beta}_m ]   =   0 \; , &
\\
      & [ \vph^{\alpha}_m , \pi^{\beta}_n ]   =
       i \hbar \Id \delta_{nm} \delta_{\alpha \beta}
        \; . &
\ea
		Here
        $ [ \;\; , \;\, ] $
		is used for commutator
\be
	[a,b] = ab - ba
		\;\; ,
\ee
	$ \hbar $
                is a Planck constant,
                which we soon shall drop,
        $ \Id $
		is as unity in algebra
        $ \ALG $.
		We can call
	$ l $
                a number of degrees of freedom
                per lattice site;
                the full number of degrees of freedom is
	$ N l $.

                Each pair of canonical variables
                is represented
                in an infinite dimensional Hilbert space,
		which can be chosen as
        $ L_2(\RR) $
		with
        $ \vph $ and $ \pi $
                being operators of multiplication
                and differentiation
\be
        \vph \; f( \xi ) = \xi f( \xi )\; ; \;\;
        \pi \; f( \xi ) =
                \frac{\hbar}{i} \frac{d}{d\xi } f( \xi )
	\;\; .
\ee

   \item        Spin variables
	$ S^\alpha_n , \; \alpha = 1, 2, 3 $
		with the relations
\be
        [ S^\alpha_m , S^\beta_n ] = i \Id \hbar
        \veps_{\alpha \beta \gamma}
                        S^\gamma_{n} \delta_{mn}  \quad ,
\ee
		where
        $ \veps_{\alpha \beta \gamma} $
		is a completely antisymmetric tensor,
        $ \veps_{123} = 1 $.

                Mathematically
                these variables define a Lie algebra
        $ \sltwo $,
		the finite dimensional representations of
                which are labeled by half--integer
	$ s = 0, \; 1/2, \; 1, \; \dots $
		and are realized in
        $ \CC^{2s+1}$.
		In the smallest nontrivial dimension
	$ ( s = 1/2 ) $
		operators
	$ S^\alpha_n $
		are represented by Pauli matrices
	$ \sigma^\alpha $
\be
	\sigma^1 = \left (
\begin{array}{cc}
        0 & 1 \\
        1 & 0
\end{array}
 \right )
 \; , \;\;
	\sigma^2 = \left (
\begin{array}{cr}
        0 & -i \\
        i &  0
\end{array}
 \right )
 \; , \;\;
	\sigma^3 = \left (
\begin{array}{cr}
        1 &  0 \\
        0 & -1
\end{array}
 \right )
\ee
		as
        $ S^\alpha_n = \hbar/2 \sigma^\alpha $.

    \item       Weyl variables.
\newline
		For one degree of freedom per lattice
		site these variables consist of a pair
	$ u_n $, $ v_n $
		with the exchange relations
\ba
	& u_m u_n = u_n u_m \; ; \;\;
		v_m v_n = v_n v_m \; ;\;\;
	u_m v_n = v_n u_m \; , \;\; m \neq n\; ; &
\nonumber
			\\
	& u_n v_n = q v_n u_n \; , &
\ea
		where
	$ q $
                is a given complex number.
                Usually it assumes the values on a circle
                and is parametrized by a real number
	$ \gamma $
\be
	q = e^{i \hbar \gamma}\;\;.
\ee

    \item
        $ q $--deformed spin variables
	$ S^\alpha_n $.
                In writing the commutation relations we shall
                drop the lattice index
	$ n $
		and present them for fixed
	$ n $
		as follows
\be
	q^{S^3} S^{\pm} = q^{\pm 1} S^{\pm} q^{S^3}\;\; ;
\label{Sq}
\ee
\be
        [ S^{+} , S^{-} ] =
                \frac{(q^{S^{3}})^2 - (q^{S^{3}})^{-2}}
                        { q - q^{-1} }           \quad ,
\label{Spm}
\ee
		via generators denoted by
        $ q^{S^{3}} $, $ S^{+} $ and $ S^{-} $.
		When
	$ q \rightarrow 1 $ (or $\gamma\rightarrow 0$)
		these relations turn into the usual
        $ \sltwo $
		relations for
	$ S^\alpha $, $S^{\pm}$
		being the usual combinations
	$ S^{\pm} = S^1 \pm i S^2 $.

		Thus they define a
	$ q $--deformed
        $ \sltwo $
		algebra, denoted by
        $ \slqtwo $.
		For generic
	$ q $
                the finite dimensional representations are
                given in the same spaces
	$ \CC^{2s+1} $
		as in nondeformed case. However for
	$ q $
                on the circle some new interesting
                representations occur, more on this below.
\end{enumerate}

                The Hilbert space for the representation of
                ultralocal algebra
        $ \ALG $
                has a natural tensor--product form
\be
	 \HH = \prod_{n=1}^N \otimes \hh_n =
		\hh_1 \otimes \hh_2 \otimes \dots \otimes
			\hh_n \otimes \dots \otimes \hh_N
\ee
		(where all
	$ \hh_n $
		could be the same) and variables
	$ X_n^\alpha $
		act nontrivially only in the space
	$ \hh_n $
\ba
        X_n^\alpha = \Id \otimes \Id \otimes \dots
                & \otimes X^\alpha \otimes &
                        \dots \otimes \Id \;\; . \\
        & \hbox{ $n$--th place} &
\nonumber
\ea
                In the continuous limit we encounter the
                problem of considering the infinite tensor
                products, which is known to be intricated from
                the time of v.Neumann. In fact, the concrete
                examples of the continuous limits will give
                the instances of the rather nontrivial
                constructions of such products.

                The simplest example of the nonultralocal
                relation is furnished by the exchange algebra
                for variables
	$ w_n $
		with the relations
\ba
	w_{n} w_{n+1} = q w_{n+1} w_{n} \, ; \;&
	w_{m} w_{n} = w_{n} w_{m} & \;\;
                        |n-m| \geqslant 2 \;\; .
\ea

                The Hilbert space for this algebra is also a
                tensor product, but with the length being
                around
	$ N/2 $.
                We shall not discuss it here, however, the
                example being given just for illustration.

                General considerations
        {\it a-l\'a}
		Darboux theorem in
                classical mechanics state that the canonical
                variables are generic in the sense that all
                other types of dynamical variables can be
                expressed through them. Let us illustrate it
                on the examples above.

		Let
	$ \psi $, $ \psi^* $
                be a complex pair of canonical variables with
                relation
\be
        [ \psi , \psi^* ] =  \Id
\ee
        (i.e. $ \psi = \frac{1}{\sqrt{2}}(\vph + i \pi) $,
                $ \psi^* = \frac{1}{\sqrt{2}}(\vph - i\pi) $);
		then the variables
\ba
        S^+ & = & S^1 + i S^2 = \psi^* ( 2s - \psi^* \psi ) \; ; \\
        S^- & = & S^{1} - i S^{2} = \psi \; ; \\
        S^3 \; & = & \psi^* \psi - s
\ea
                satisfy the spin commutation relation.
                Here
	$ s $
                is any complex parameter,
                which can be called spin, as the Casimir
	$ C = (S^1)^2 + (S^2)^2 + (S^3)^2 $
		has value
	$ C = s(s+1) $.

		Weyl pair
	$ u $, $ v $
		can be realized via canonical variables
        $ \pi $ and $ \vph $
		as follows
\be
        u = e^{i \pi Q} \; , \;\; v = e^{i \vph P} \;\; ,
\label{WP}
\ee
		where
	$ P $ and $ Q $
		are any complex numbers and
	$ q = e^{iPQ} $.

		Finally, the deformed spin variables
	$ q^{S^3}$, $ S^{\pm} $
		can be realized as
\ba
	& S^{\pm} = e^{\pm i\pi/2}
                ( 1 + m^2 e^{\pm 2i \gamma \vph})
                        e^{\pm i\pi/2} &  ;
\label{qSpins} \\
       & q^{S^3} = e^{i \gamma \vph} &  ,
\label{qSpin}
\ea
		where
	$ m^2 $ and $ \gamma $
		are parameters.

                Of course the Hilbert space for the canonical
                variables is infinite dimensional and the
                representations for the spin variables appear
                as reductions.
                For example, if
	$ 2s $
		is integer then the subspace consisting of
	$ 2s+1 $
		states
	$ \omega $, $ \psi^* \omega $, \dots
                $ (\psi^*)^{2s} \omega $,
		where
	$ \omega $
		is a vector annihilated by
	$ \psi $, $ \psi \omega = 0 $,
		is invariant with respect to action of
	$ S^{\pm} $, $ S^{3} $.

		The Weyl pair
	$ u $, $ v $
		in form
	(\ref{WP})
		has finite dimensional representation if
	$ q $
		is a root of unity, or when
	$ PQ/2\pi $
                is a rational number.
                Corresponding reduction exists also for the
                deformed spin variables if
	$ \gamma / 2\pi $
                is rational and it gives the so-called finite
                dimensional cyclic representation of
        $ \slqtwo $.

                These reductions evidently reflect some global
                aspects of the would--to--be quantum Darboux
                theorem.

                This completes my short introduction to the
                kinematics of the models I plan to consider.
                I come to the representative example --- spin
	$ 1/2 $
		XXX chain.

\section{ XXX${}_{1/2}$ model. Description. }
\label{sthree}

                The origin of the abbreviation XXX will become
                clear soon. As I have already said this is
                quantum model, defined in a Hilbert space
\be
	 \HH_N = \prod_{n=1}^N \otimes \hh_n \;\; ,
\ee
		where each local space
	$ \hh_n $
                is two--dimensional
\be
	\hh_n = \CC^2 \;\;\; .
\ee
		The spin variables
	$ S_n^\alpha $
		are acting on each
	$ \hh_n $
                as Pauli matrices divided by $ 2 $.

                There are several important observables, such
                as the total spin
\be
	S^\alpha = \sum_n S_n^\alpha
\ee
                or the total hamiltonian
\be
	\HAM = \sum_{\alpha,n} \left( S_n^\alpha
			S_{n+1}^\alpha - \frac{1}{4} \right)
				\;\; ,
\label{Hamhalf}
\ee
		where the periodicity
\be
	S_{n+N}^\alpha  = S_{n}^\alpha
\ee
		is to be taken into account. We have
\be
	\left[ \HAM , S^\alpha \right] = 0 \;\; ,
\ee
		which reflects the
        $ \sltwo $
		symmetry of the model.

                The abbreviation XXX is used to stress this
                invariance which is reflected in the fact that
                all coefficients in front of combination
	$ S_n^\alpha S_{n+1}^\alpha $
                in hamiltonian are equal.
                A more general hamiltonian
\be
	\HAM = \sum_{\alpha,n} J^\alpha
			S_n^\alpha S_{n+1}^\alpha
\ee
		with parameters
	$ J^\alpha $
		corresponds to XYZ spin
	$ 1/2 $
		model.

		Our problem is to investigate the spectrum of
	$ \HAM $.
		Of course for finite
	$ N $
		it is just a problem about the matrix
	$ 2^N \times 2^N $
                accessible by computer.
                However we will be interested in the limit
	$ N \rightarrow \infty $,
		when only analytic methods work.

                The core of our approach is a generating
                object, called Lax operator.
                This object is a rather long shot from the Lax
                operator of KdV equation but historically this
                name was fixed for any linear operator,
                entering the auxiliary spectral problem of the
                classical inverse scattering method.

                The definition of the Lax operator involves
                the local quantum space
	$ \hh_n $
		and the auxiliary space
	$ \VV $,
		which for the beginning will be also
	$ \CC^2 $.
		Lax operator
	$ L_{n,a}(\lambda) $
		acts in
	$ \hh_n \otimes \VV $
		and is given explicitly by the expression
\be
        L_{n,a}(\lam) = \lam \Id_n \otimes \Id_a +
		i \sum_\alpha S_n^\alpha \otimes
			\sigma^\alpha \;\; ,
\ee
		where
        $ \Id_n $, $ S_n^\alpha $
		act in
	$ \hh_n $ and
        $ \Id_a $, $ \sigma^\alpha $
		are unit and Pauli matrices in
        $ \VV = \CC^2 $; $ \lam $
                is a complex parameter, usually called the
                spectral parameter, reminding its role as an
                eigenvalue in the original Lax operator.

		Alternatively
	$ L_{n,a}(\lambda) $
		can be written as
	$ 2 \times 2 $
		matrix
\be
	L_{n,a}(\lambda) = \left (
		\begin{array}{cc}
        \lam + i S_n^3 & i S_n^- \\
        i S_n^+       & \lam - i S_n^3
		\end{array} \right ) \;\; ,
\label{Laxlam}
\ee
		acting in
	$ \VV $
		with entries being operators in quantum space
	$ \hh_n $.
		One more form uses the fact that operator
\be
        \PP = \frac{1}{2} ( \Id \otimes \Id +
                 \sum_\alpha \sigma^\alpha \otimes
			\sigma^\alpha )
\label{PII}
\ee
		is a permutation in
	$ \CC^2 \otimes \CC^2 $
\be
	\PP \; a \otimes b = b \otimes a \;\;\; .
\label{Permut}
\ee
		In terms of
	$ \PP_{n,a} $,
                which makes sense due to the fact that
	$ \hh_n $ and $ \VV $
		are the same
	$ \CC^2 $,
		we have
\be
        L_{n,a}(\lam) = ( \lam - \frac{i}{2} ) \Id_{n,a} +
		i \PP_{n,a} \;\; .
\label{Laxtwotwo}
\ee

%


                Now we establish the main property of Lax
                operator --- the commutation relation for its
                entries. As we have four of them we are to
                write down 16 relations. Our convenient
                notations allow to write them all in one line.

                Consider two exemplars of Lax operators
	$ L_{n,a_1}(\lambda) $ and
	$ L_{n,a_2}(\mu) $
		with the same quan\-tum space and
	$ \VV_1 $ and $ \VV_2 $
                ser\-ving as cor\-res\-pon\-ding auxi\-liar
                spaces.
                The products
        $ L_{n,a_1}(\lam) \; L_{n,a_2}(\mu) $ and
        $ L_{n,a_2}(\mu) \; L_{n,a_1}(\lam) $
                make sense in a triple tensor product
	$ \hh_n \otimes \VV_1 \otimes \VV_2 $.
                We claim, that these two products are similar
                operators with the intertwiner acting only in
	$ \VV_1 \otimes \VV_2 $
                and so not containing quantum operators.
                In other words, there exists an operator
        $ R_{a_1,a_2}( \lam - \mu ) $ in
			$ \VV_1 \otimes \VV_2 $
		such that the following relation is true
\be
 	R_{a_1,a_2}( \lam - \mu ) \,
                L_{n,a_1}(\lam) \, L_{n,a_2}(\mu) \,=\,
              L_{n,a_2}(\mu) \, L_{n,a_1}(\lam)
                \,      R_{a_1,a_2}( \lam - \mu )\;.
\label{FCR}
\ee
		The explicit expression for
        $ R_{a_1,a_2}( \lam ) $ is
\be
        R_{a_1,a_2}( \lam )  = \lam \Id_{a_1,a_2} +
		i \PP_{a_1,a_2} \;\; ,
\label{Rtwotwo}
\ee
        where $ \Id_{a_1,a_2} $ and $ \PP_{a_1,a_2} $
		are unity and permutation in
	$ \VV_1 \otimes \VV_2 $.

		Comparing
	(\ref{Laxtwotwo}) and
	(\ref{Rtwotwo})
		we see that the Lax operator
        $ L_{n,a}(\lam) $
		and operator
        $ R_{a_1,a_2}( \lam ) $,
		which we shall call
	$ R $--matrix,
		are essentially the same.

		To check
	(\ref{FCR})
		it is convenient to use the form
	(\ref{Rtwotwo})
		for
        $ L_{n,a}(\lam) $
                and the commutation relation for permutations
\be
	\PP_{n,a_1} \, \PP_{n,a_2} \, = \,
		\PP_{a_1,a_2} \, \PP_{n,a_1} \, = \,
			\PP_{n,a_2} \, \PP_{a_2,a_1}
\label{Pna}
\ee
		together with evident symmetry
\be
	\PP_{a_2,a_1} \; = \; \PP_{a_1,a_2} \;\;.
\label{Paa}
\ee

		The importance of the relation
	(\ref{FCR})
                will become clear momentarily.
                We shall call it in what follows the
                fundamental commutation relation (FCR).
                Its general place in the family of the
                Yang--Baxter relations will become evident
                later.

                The Lax operator
        $ L_{n,a}(\lam) $
                has a natural geometric interpretation as a
                connection along our chain, defining the
                transport between sites
	$ n $ and $ n+1 $
		via the Lax equation
\be
	\psi_{n+1} \, = \, L_n \psi_{n}
\ee
		for vector
	$ \psi_n = \left (
		\begin{array}{c} \psi_n^1 \\
				  \psi_n^2
		\end{array} \right ) $
		with entries in
	$ \HH $.
		The ordered product over all sites between
	$ n_2 $ and $ n_1 $
\be
        T_{n_1,a}^{n_2}(\lam) =
                L_{n_2,a}(\lam) \, \dots \, L_{n_1,a}(\lam)
\ee
		defines the transport form
	$ n_1 $ to $ n_2 +1 $
		and the full product
\be
        T_{N,a}(\lam) \, = \,
                L_{N,a}(\lam) \, \dots \, L_{1,a}(\lam)
\ee
                is a monodromy around our circle.
                The last operator is given as a
	$ 2 \times 2 $
		matrix in the auxiliary space
	$ \VV $
\be
	T_{N,a} = \left (
                \begin{array}{ccc} A_N(\lam) & , & B_N(\lam)
                        \\  C_N(\lam) & , & D_N(\lam)
		\end{array} \right )
\ee
                with entries being operators in the full
                quantum space
	$ \HH $.
                As in classical case the map from local
                dynamical variables
	( $ S_n^\alpha $ )
		to monodromy
        $ T_{N,a}(\lam) $
                is a tool for solving the dynamical problem.
                We shall see that
        $ T_{N,a}(\lam) $
                is a generating object for main observables
                such as spin and hamiltonian, as well as for
                the spectrum rising operators.

		For that we shall establish first the FCR for
        $ T_{N,a}(\lam) $.
                We claim, that it has exactly the same form as
                the local FCR
	(\ref{FCR}),
		namely
\be
        R_{a_1,a_2}( \lam - \mu ) \,
                T_{a_1}(\lam) \, T_{a_2}(\mu) \,=\,
        T_{a_2}(\mu) \, T_{a_1}(\lam)
                \,      R_{a_1,a_2}( \lam - \mu ) \; .
\label{RTT}
\ee
		We dropped here the index
	$ N $
                and shall do it in what follows as soon as it
                does not lead to confusion.

		Derivation of
	(\ref{RTT})
                is very simple and uses the advantages of our
                notations.
                We shall prove it for any transport
                operator.
                It is clear, that it is enough to consider the
                transport along two sites, i.e.
	$ n \rightarrow n+2 $.
		With short notations
        $ R_{a_1,a_2}( \lam - \mu ) = R_{12} $,
        $ L_{n,a_1}(\lam) = L_{1} $,
        $ L_{n+1,a_1}(\lam) = L'_{1} $,
	$ L_{n,a_2}(\mu) = L_{2} $,
	$ L_{n+1,a_2}(\mu) = L'_{2} $
		we have
\begin{eqnarray}
       	R_{12}  L'_{1}  L_{1}  L'_{2}  L_{2}
   & = &
        \hbox{(due to commutativity of
$   L_{1}  $ and $ L'_{2}  $) }
\nonumber
\\
      =  R_{12}  L'_{1}  L'_{2}  L_{1}  L_{2}
   & = &
        \hbox{(due to the local FCR for
$   L_{1}  $, $ L_{2}  $ and $  L'_{1}  $, $ L'_{2}  $) }
\nonumber
\\
      =   L'_{2}  L'_{1}  L_{2}  L_{1} R_{12}
   & = &
        \hbox{(due to commutativity of
$  L'_{1}  $ and $ L_{2}  $) }
\nonumber
\\
      =   L'_{2}  L_{2}  L'_{1}  L_{1} R_{12}
   & \; . &
\nonumber
\end{eqnarray}
		This completes the proof.

		The monodromy
        $ T_{N,a}(\lam) $
		is a polynomial in
        $ \lam $
		of order
	$ N $
\be
        T_{N,a}(\lam) = \lam^{N} + i \lam^{N-1} \sum_{\al}
		(S^\alpha \otimes \sigma^\alpha)
			+ \ldots \;\; ,
\label{TNa}
\ee
                so that total spin
	$ S^\alpha $
                appears via the coefficient of next to the
                highest degree. Now we shall find the place
                for the hamiltonian. The FCR
	(\ref{RTT})
		shows, that the family of operators
\be
        F(\lam) = \tr \, T(\lam) = A(\lam) + D(\lam)
\ee
		is commuting
\be
        [ F(\lam) , F(\mu) ] = 0 \;\; .
\ee
		Its nontrivial
        $ \lam $
                expansion begins with power
        $ \lam^{N-2}$
\be
        F(\lam) = 2 \lam^{N} + \sum_{l=0}^{N-2}
                        Q_l \lam^{l}
\ee
		and produces
	$ N-1 $
		commuting operators
	$ Q_l $.
		We shall show, that
	$ \HAM $
		belongs to this family.

		For this note, that the point
        $ \lam = i/2 $
		is rather special
\be
        L_{n,a} (i/2)
			= i \PP_{n,a}
\ee
		and of course for any
        $ \lam $
\be
        \frac{d}{d\lam} L_{n,a} (\lam) = \Id_{n,a} \;\; .
\ee
		This makes it easy to control the expansion of
        $ F(\lam) $
		in  the vicinity of
        $ \lam = i/2 $.

		We have
\be
        T_{N,a}(i/2) = i^N
		\PP_{N,a} \, \PP_{N-1,a} \,
                         \dots \, \PP_{1,a} \quad .
\ee
                This string of permutations is easily
                transformed into
\be
		\PP_{1,2} \, \PP_{2,3} \, \dots \;
		\; \PP_{N-1,N} \, \PP_{N,a}
\ee
                by taking permutations one after another from
                left to right and taking into account the
                properties
	(\ref{Pna}) and
	(\ref{Paa})
		and commutativity of
                permutations with completely different indeces.
                Now the trace over the auxiliary space is
                easily taken
\be
        \tr_{a}\PP_{N,a} \, = \, \Id_{N}   \quad ,
\ee
		so that
\be
        U = i^{-N} \tr_a T_{N} (i/2) =
		\PP_{1,2} \, \PP_{2,3} \, \dots \;
		\; \PP_{N-1,N} \;\; .
\ee
		It is easy to see, that
	$ U $
		is a shift operator in
	$ \HH $.
		The property
	(\ref{Permut})
		can be rewritten as
\be
	\PP_{n_1,n_2} \, X_{n_2} \, \PP_{n_1,n_2} \, =
			\, X_{n_1} \;\; .
\ee
		Thus
\ba
	X_n U =
		\PP_{1,2} \dots X_n \,
			\PP_{n-1,n} \, \PP_{n,n+1}
		\dots \PP_{N-1,N} \, =
\nonumber
	\\
		\PP_{1,2} \dots
			\PP_{n-1,n}\, X_{n-1}\, \PP_{n,n+1}
		\dots \PP_{N-1,N} =
	U X_{n-1} \;\; .
\ea
		Operator
	$ U $
		is unitary
\be
        U^* \, U \, = \, U \, U^* \,= \, \Id \;\; ,
\ee
		because the permutations have properties
\be
	\PP^* \, = \, \PP \, ;
                \,\,  \PP^2 \,= \, \Id \;\; ,
\ee
		and we have
\be
	U^{-1} \, X_n \, U \, = \, X_{n-1} \;\; ,
\ee
                which allows to introduce important
                observable --- momentum. By definition,
                momentum
	$ \PMO $
                produces an infinitesimal shift and on the
                lattice it is substituted by shift along one
                site
\be
	e^{i \PMO} \, = \, U \;\; .
\ee

		We proceed now to expand
        $ F(\lam) $
		in the vicinity of
        $ \lam = i/2 $.
		We get
\be
	\left.
                \frac{d}{d\lam} T_a(\lam)
        \right|_{\lam = i/2} \, = \,
		i^{N-1} \sum_n
		\PP_{N,a} \ldots
			\widehat{\PP}_{n,a}
		\ldots \PP_{1,a} \; ,
\ee
		where
	$ \; \widehat{} \; $
                means that corresponding factor is absent.
                Repeating our trick we transform this
                after taking the trace over
	$ \VV $ into
\be
	\left.
                \frac{d}{d\lam} F_a(\lam)
        \right|_{\lam = i/2} \, = \, i^{N-1} \sum_n
		\PP_{1,2} \ldots \PP_{n-1,n+1}
		\ldots \PP_{N-1,N} \;\; .
\ee
                We can cancel most of permutations here,
                multiplying by
	$ U^{-1} $;
		as a result we get
\be
	\left.
                \frac{d}{d\lam} F_a(\lam) \; F_a(\lam)^{-1}
        \right|_{\lam = i/2} =
	\left.
                \frac{d}{d\lam} \ln F_a(\lam)
        \right|_{\lam = i/2} =
                \frac{1}{i} \sum_n \PP_{n,n+1} \; .
\label{FaFa}
\ee
		Using
	(\ref{PII})
		we can rewrite the expression
	(\ref{Hamhalf})
		for the hamiltonian as
\be
        \HAM \, = \, \frac{1}{2} \sum_n \PP_{n,n+1}
                        \, - \, \frac{N}{2}
\label{HPN}
\ee
		and comparing
	(\ref{HPN})
		and
	(\ref{FaFa})
		we have
\be
	\HAM \, = \, \left.
                \frac{i}{2} \frac{d}{d\lam} \ln F(\lam)
                     \right|_{\lam = i/2}
                        \, - \, \frac{N}{2} \;\; .
\ee
		Thus I have shown, that
	$ \HAM $
		indeed belongs to the family of
	$ N - 1 $
                commuting operators generated by the trace of
                monodromy
        $ F(\lam) $.
		One component of spin, say
	$ S^3 $,
		completes this family to
	$ N $
                commuting operators. This is a proof of the
                integrability of the classical counterpart of
                our model, which can be considered as a system
                with
	$ N $
                degrees of freedom. In the next section I
                shall show, how to describe the set of
                eigenstates of commuting family
        $ F(\lam) $
		using the offdiagonal elements of monodromy.

\section{XXX${}_{1/2}$ model. Bethe Ansatz equations.}
\label{sfour}

                Here I shall describe a procedure to
                diagonalize the whole family of operators
        $ F(\lam) $
                in a rather algebraic fashion, based on the
                global FCR
	(\ref{RTT})
                and some simple properties of local Lax
                operators. In a way the working
                generalizes the simplest quantum mechanical
                treatment of harmonic oscillator hamiltonian
	$ n = \psi^* \psi $
		based on commutation relations
        $ [ \psi , \psi^* ] = \Id $
		and existence of a state
	$ \omega $
		such that
	$ \psi \, \omega = 0 $.

		Let us write the relevant set of FCR
\ba
        & [ B(\lam) , B(\mu) ] \, = \, 0 \;\; ; & \\
        & A(\lam) \, B(\mu) \, = \,
                f(\lam-\mu) \, B(\mu) \, A(\lam) \, + \,
                g(\lam-\mu) \, B(\lam) \, A(\mu) & ;
\label{ExchAB} \\
        & D(\lam) \, B(\mu) \, = \,
                h(\lam-\mu) \, B(\mu) \, D(\lam) \, + \,
                k(\lam-\mu) \, B(\lam) \, D(\mu) & ,
\label{ExchDB}
\ea
		where
\ba
        f(\lam) \, = \, \frac{\lam - i}{\lam} &;\;\;\;\;\; &
        g(\lam) \, = \, \frac{i}{\lam} \; ;
\nonumber \\
        h(\lam) \, = \, \frac{\lam + i}{\lam} &;\;\;\;\;\; &
        k(\lam) \, = \, - \frac{i}{\lam} \; .
\label{fghk}
\ea
		To read these relations from one line FCR
	(\ref{RTT})
                one must use the explicit matrix
		representation of
                FCR in
	$ \VV \otimes \VV $.
		All objects in it are
	$ 4 \times 4 $
		matrices in a natural basis
\be
	e_1 = e_+ \otimes e_+ \, , \;
	e_2 = e_+ \otimes e_- \, , \;
	e_3 = e_- \otimes e_+ \, , \;
	e_4 = e_- \otimes e_-
\ee
		in
	$ \CC^2 \otimes \CC^2 $, where
\be
	e_+ =
		\left(
	\begin{array}{c} 1 \\ 0 \end{array}
		\right) \, , \;\;\;\;
	e_- =
		\left(
	\begin{array}{c} 0 \\ 1 \end{array}
		\right) \; .
\ee
                Matrix
	$ \PP $
                assumes the form
\be
	\PP =
		\left( \begin{array}{cccc}
	1 & 0 & 0 & 0 \\
	0 & 0 & 1 & 0 \\
	0 & 1 & 0 & 0 \\
	0 & 0 & 0 & 1
			\end{array} \right) \;\; ,
\ee
		so that
        $ R(\lam) $
		looks like
\be
        R(\lam) =
		\left( \begin{array}{cccc}
        a(\lam) &         &         &     \\
               & b(\lam)  & c(\lam)  &     \\
               & c(\lam)  & b(\lam)  &     \\
               &         &         & a(\lam)
			\end{array} \right)
\label{Rfour}
\ee
		(we do not write in the zeros), where
\be
        a \, = \, \lam + i \, ,\;\;
        b \, = \, \lam \, ,\;\;
	c \, = \, i \;\; .
\ee
		The matrices
        $ T_{a_1}(\lam) $ and $ T_{a_2}(\mu) $
		take the form
\be
        T_{a_1}(\lam) =
		\left( \begin{array}{cccc}
        A(\lam) &         & B(\lam)  &        \\
               & A(\lam)  &         & B(\lam) \\
        C(\lam) &         & D(\lam)  &        \\
               & C(\lam)  &         & D(\lam)
			\end{array} \right)
\ee
		and
\be
        T_{a_2}(\mu) =
		\left( \begin{array}{cccc}
        A(\mu) & B(\mu)  &         &        \\
        C(\mu) & D(\mu)  &         &        \\
                &        &  A(\mu) & B(\mu) \\
                &        &  C(\mu) & D(\mu)
			\end{array} \right) \;\; .
\ee
		Thus we have
\be
        T_{a_1}(\lam) \, T_{a_2}(\mu) =
		\left( \begin{array}{ccccc}
A(\lam)A(\mu) &A(\lam)B(\mu) & &B(\lam)A(\mu) &B(\lam)B(\mu) \\
A(\lam)C(\mu) &A(\lam)D(\mu) & &B(\lam)C(\mu) &B(\lam)D(\mu) \\
C(\lam)A(\mu) &C(\lam)B(\mu) & &D(\lam)A(\mu) &D(\lam)B(\mu) \\
C(\lam)C(\mu) &C(\lam)D(\mu) & &D(\lam)C(\mu) &D(\lam)D(\mu) \\
			\end{array} \right)
\ee
		and
        $ T_{a_2}(\mu) \, T_{a_1}(\lam) $
                is given by matrix, where in all matrix
                elements the factors have opposite order.
                Now to get
	(\ref{ExchAB})
		one is to use the
	$ (1,3) $
		relation in FCR
	(\ref{RTT})
\be
        a(\lam - \mu) \, B(\lam) \, A(\mu) \, = \,
                c(\lam-\mu) \, B(\mu) \, A(\lam) \, + \,
                b(\lam-\mu) \, A(\mu) \, B(\lam) \\
\ee
		and interchange
        $ \lam \, \leftrightarrow \, \mu $
		to get
\be
        A(\lam) \, B(\mu) \, = \,
          \frac{a(\mu-\lam)}{b(\mu-\lam)}
                \, B(\mu) \, A(\lam) \, - \,
          \frac{c(\mu-\lam)}{b(\mu-\lam)}
                \, B(\lam) \, A(\mu) \; .
\ee
		Other relations are obtained similarly.

		The exchange relations
	(\ref{ExchAB})
		and
	(\ref{ExchDB})
		substitute the relations
\be
	\psi \, n \, = \, (n+1) \psi \, , \;\;
	\psi^* \, n \, = \, (n-1) \, \psi^*
\ee
		for the harmonic oscillator with
	$ n \, = \, \psi^* \psi $.
                Now the analogue of ``highest weight''
	$ \omega $
		such as
	$ \psi \, \omega \, = \, 0 $
		will be played by a reference state
	$ \Omega $
		such, that
\be
        C(\lam) \, \Omega \, = \, 0 \;\;\; .
\ee
		To find this state we observe that in each
	$ \hh_n $
		there exists a vector
	$ \omega_n $
		such that the Lax operator
        $ L_{n,a}(\lam) $
                becomes triangular in the auxiliary space,
                when applied to it
\be
        L_{n}(\lam) \, \omega_n \, = \,
		\left( \begin{array}{cc}
                        \lam + \frac{i}{2} & *  \\
                         0 & \lam - \frac{i}{2}
		\end{array} \right)
                         \; \omega_n  \quad .
\ee
		This vector is given by
	$ \omega_n = e_+ $.
		By
	$ * $
                we denote operator expressions which are
                not relevant for us. Now for vector
	$ \Omega $ in
	$ \HH $
\be
	\Omega \, = \, \prod_n \otimes \, \omega_n
\ee
		we get
\be
        T(\lam) \, \Omega \, = \,
		\left( \begin{array}{cc}
                        \alpha^N(\lam) & *  \\
                         0 & \delta^N(\lam)
		\end{array} \right)
			 \, \Omega \;\; ,
\ee
		where
\be
        \alpha(\lam) \, = \, \lam + \frac{i}{2} \, , \quad
        \delta(\lam) \, = \, \lam - \frac{i}{2} \, .
\label{aldel}
\ee
		In other words we have
\be
        C(\lam) \, \Omega \, = \, 0 \, ; \;\;
        A(\lam) \, \Omega \, =
                \, \alpha^N(\lam) \, \Omega \, ; \;\;
        D(\lam) \, \Omega \, = \, \delta^N(\lam) \, \Omega\; ,
\ee
		so that
	$ \Omega $
		is an eigenstate of
        $ A(\lam) $ and  $ D(\lam) $
                simultaneously and also that for
	$ F \, = \, A \, + \, D \, $.

		Other eigenvectors will be looked for
		in the form
\be
        \Phi(\{\lam\}) \, = \,
                B(\lam_1) \, \dots \, B(\lam_l) \,\Omega\; .
\ee
		The condition that
        $ \Phi(\{\lam\}) $
		is an eigenvector of
        $ F(\lam) $
                will lead to a set of algebraic relations on
                parameters
        $ \lam_1 $, \dots $ \lam_l $.
		I proceed to derive these equations.

		Using the exchange relations
	(\ref{ExchAB})
		we get
\ba
     && A(\lam) \, B(\lam_1) \, \dots \, B(\lam_l) \, \Omega
		\, = \, \prod_{k=1}^{l} \,
                        f(\lam-\lam_k) \, \alpha^N(\lam) \,
                B(\lam_1) \, \dots \, B(\lam_l) \, \Omega +
\nonumber \\
              && + \, \sum_{k=1}^{l} \, M_k(\lam,\{\lam\}) \,
        B(\lam_1) \, \dots \, \widehat{B}(\lam_k) \, \dots \,
                        B(\lam_l) \, B(\lam) \, \Omega \;\; .
\label{ABBO}
\ea
                The first term in RHS has a desirable form and
                is obtained using only the first term in the
                RHS of
	(\ref{ExchAB}).
		All other terms are combinations of
	$ 2^l-1 $
		terms, which one has taking
        $ A(\lam) $
		to
	$ \Omega $
		using the exchange relation
	(\ref{ExchAB}).
		The coefficients
	$ M_k $
                can be quite involved. However the
                coefficient
	$ M_1 $
		is simple enough, to get it one
                must use the second term in
	(\ref{ExchAB})
		during the interchange of
        $ A(\lam) $ and $ B(\lam_1) $
                and in all other exchanges use only the first
                term in RHS of
	(\ref{ExchAB}).
		Thus we get
\be
        M_1(\lam,\{\lam\}) \, = \, g(\lam-\lam_1) \,
		\prod_{k=2}^{l} \,
                        f(\lam_1-\lam_k) \, \alpha^N(\lam_1)\; .
\ee
                Now we comment, that due to the commutativity
                of
        $ B(\lam) $
		all other coefficients
        $ M_j(\lam,\{\lam\}) $
		are obtained from
        $ M_1(\lam,\{\lam\}) $
		by a simple substitution
        $ \lam_1 \, \to \, \lam_j $
		so that
\be
        M_j(\lam,\{\lam\}) \, = \, g(\lam-\lam_j) \,
		\prod_{k\neq j}^{l} \,
                f(\lam_j-\lam_k) \, \alpha^N(\lam_j) \; .
\ee

		This means, of course, that the coefficients
        $ a(\lam) $, $ b(\lam) $, $ c(\lam) $,
		entering
	$ R $--matrix,
                satisfy involved sum rules, making the FCR
                consistent.

		Analogously for
        $ D(\lam) $
		we have
\ba
    && D(\lam) \, B(\lam_1) \, \dots \, B(\lam_l) \, \Omega
		\, = \, \prod_{k=1}^{l} \,
                        h(\lam-\lam_k) \, \delta^N(\lam) \,
                B(\lam_1) \, \dots \, B(\lam_l) \, \Omega +
\nonumber \\
             && + \, \sum_{k=1}^{l} \, N_k(\lam,\{\lam\}) \,
       B(\lam_1) \, \dots \, \widehat{B}(\lam_k) \, \dots \,
                        B(\lam_l) \, B(\lam) \, \Omega \;\; ,
\label{DBBO}
\ea
		where
\be
        N_j(\lam,\{\lam\}) \, = \, k(\lam-\lam_j) \,
		\prod_{k\neq j}^{l} \,
                h(\lam_j-\lam_k) \, \delta^N(\lam_j) \; .
\ee

		Observe now, that
\be
        g(\lam-\lam_j) \, = \, -k(\lam-\lam_j) \;\; .
\ee
		This allows to cancel the unwanted terms in
	(\ref{ABBO})
		and
	(\ref{DBBO})
		for the application of
        $ A(\lam) \, + \,  D(\lam) $ to $ \Phi(\{\lam\}) $.
		We get, that
\be
        (A(\lam)\, + \,D(\lam)\,) \, \Phi(\{\lam\}) \, = \,
                \Lambda(\lam,\{\lam\})\, \Phi(\{\lam\})
\ee
		with
\be
        \Lambda(\lam,\{\lam\})\, = \,
                \alpha^N(\lam) \prod_{j=1}^{l}
                        f(\lam-\lam_j) \, + \,
                \delta^N(\lam) \prod_{j=1}^{l}
                        h(\lam-\lam_j) \; ,
\ee
		if the set of
        $ \{\lam\} $
		satisfy the equations
\be
	\prod_{k\neq j}^{l} \,
                f(\lam_j-\lam_k) \, \alpha^N(\lam_j) \, = \,
	\prod_{k\neq j}^{l} \,
                h(\lam_j-\lam_k) \, \delta^N(\lam_j)
\label{falh}
\ee
		for
	$ j\,=\,1,\ldots,l $.
		Using the explicit expressions
	(\ref{fghk})
		and
	(\ref{aldel})
		we rewrite
	(\ref{falh})
		in the form
\be
	\left(
                \frac{\lam_j+i/2}{\lam_j-i/2}
	\right)^N \, = \,
		\prod_{k\neq j}^{l} \,
        \frac{\lam_j-\lam_k+i}{\lam_j-\lam_k-i} \quad .
\label{BAE}
\ee
                This is the main result of this section.
                In what follows we shall use the equations
	(\ref{BAE})
		to investigate the
	$ N \, \to \, \infty $
		limit.

		An important observation is, that equations
	(\ref{BAE})
                mean, that the superficial poles in the
                eigenvalue
        $ \Lambda(\lam,\{\lam\}) $
		actually cancel so that
        $ \Lambda $
		is a polynomial in
        $ \lam $
		of degree
	$ N $
                as it should. This observation makes one
                think that only solutions
        $ \{\lam\} $
		with
        $ \lam_j \neq \lam_k $
		are relevant for our purpose. Indeed, equal
        $ \lam $
                will lead to higher order spurious poles,
                cancelling of which requires more than
	$ l $
                equations. And indeed we shall see below, that
                solutions with nonequal
        $ \lam_j $
		are enough to give all spectrum.

		The equations
	(\ref{BAE})
                appeared first (in a different form) in the
                paper of H.Bethe in 1931, in which exactly the
                hamiltonian
	$ \HAM $
                was investigated. The algebraic derivation in
                this lecture is completely different from the
                original approach of Bethe, who used an
                explicit Ansatz for the eigenvectors
	$ \Phi $
                in a concrete coordinate representation for
                the spin operators. The term Bethe Ansatz
                originates from that paper. We propose to
                call our approach ``The Algebraic Bethe Ansatz''
                (ABA). The equations
	(\ref{BAE})
                and vector
        $ \Phi(\{\lam\}) $
                will be called Bethe Ansatz equations (BAE)
                and Bethe vector correspondingly.

                We finish this section by giving the explicit
                expressions for the eigenvalues of the
                important observables on Bethe vectors. We
                begin with the spin.

		Taking limit
	$ \mu \, \to \, \infty $
		in FCR
	(\ref{RTT})
		and using
	(\ref{TNa})
		we get the following relation
\be
	\left[ \, T_a(\lam) \, , \, \frac{1}{2}
		\sigma^{\alpha} + S^\alpha \,
	\right] \, = \, 0 \;\; ,
\ee
		which expresses the
        $ \sltwo $
                invariance of the monodromy in the combined
                space
	$ \HH \otimes \VV $.
		From here
		we have in particular
\ba
	& \left[ \, S^3 \, , \, B \,\right] \, = \, -B & ,
\label{SThreeB} \\
	& \left[ \, S^+ \, , \, B \,\right] \, = \,
			A \, - \, D  &.
\label{SPlusB}
\ea
		Now for reference state
	$ \Omega $
		we have
\be
	S^+ \, \Omega \, = \, 0 \, , \;\;\;
        S^3 \, \Omega \, = \, \frac{N}{2} \, \Omega \;\; ,
\label{SPlusO}
\ee
                showing, that it is the highest weight for
                spin
	$ S^\alpha $.

		From
	(\ref{SThreeB})
		and
	(\ref{SPlusO})
		we have
\be
        S^3 \Phi(\{\lam\}) \, = \,
                \left( \frac{N}{2} \, - \, l \right)
                 \Phi(\{\lam\}) \;\; .
\ee
		Let us show, that
\be
        S^+ \, \Phi(\{\lam\}) \, = \, 0 \;\; .
\ee
		From
	(\ref{SPlusB})
		we have
\ba
     && S^+\Phi(\{\lam\}) = \sum_j
           B(\lam_1) \ldots B(\lam_{j-1})
                        (A(\lam_j)-D(\lam_{j}))
                B(\lam_{j+1}) \ldots B(\lam_l) \Omega
\nonumber \\
     &&         = \, \sum_{k} O_k(\{\lam\})
        B(\lam_1) \dots \widehat{B}(\lam_k) \dots
                        B(\lam_l) \Omega
\ea
                and repeating the procedure that was used to
                derive BAE we can show, that all coefficients
        $ O_k(\{\lam\}) $
		vanish if BAE are satisfied.

		Thus
        $ \Phi(\{\lam\}) $
		are all highest weights.
		In particular it means, that the
	$ l $
		cannot be too large, because the
	$ S^3 $
                eigenvalue of the highest weight is
                nonnegative. More exactly we have an estimate
\be
        l \; \leqslant \; \frac{N}{2} \;\;\; .
\ee
		We see that the cases of even and odd
	$ N $
		are quite different. When
	$ N $
                is even, the spin of all states is integer and
                there are
        $ \sltwo $
		invariant states, corresponding to
	$ l = N/2 $.
		For odd
	$ N $
                spins are half--integer.

		Now we turn to the shift operator. For
        $ \lam \, = \, i/2 $
                the second term (and many of its derivatives
        over $ \lambda $ ) in
        $ \Lambda(\lam,\{\lam\}) $
                vanishes and this eigenvalue becomes
                multiplicative. In particular
\be
        U \, \Phi(\{\lam\}) \, = \,
                i^N F(i/2) \Phi(\{\lam\})
			\, = \,
        \prod_j \frac{\lam_j+i/2}{\lam_j-i/2}
                \Phi(\{\lam\}) \; .
\ee
		Taking
	$ \log $
                here we see, that the eigenvalues of the
                momentum
	$ \PMO $
		are additive and
\be
        \PMO \, \Phi(\{\lam\}) \, = \,
                \sum_j \, \pp(\lam_j) \, \Phi(\{\lam\}) \;\; ,
\label{eigmom}
\ee
		where
\be
        \pp(\lam) \, = \, \frac{1}{i}
                \, \ln \frac{\lam+i/2}{\lam-i/2} \;\;\; .
\label{plambda}
\ee
                The additivity property holds also for the
                energy
	$ \HAM $.
		Differentiating
        $ \ln\Lambda $
		over
        $ \lam $
		once and putting
        $ \lam \, = \, i/2 $
		we get
\be
        \HAM \, \Phi(\{\lam\}) \, = \,
           \sum_j \,\epsilon(\lam_j)\, \Phi(\{\lam\}) \;\; ,
\ee
		where
\be
        \epsilon(\lam) \, = \, - \, \frac{1}{2} \,
                \frac{1}{\lam^2 + 1/4} \;\;\; .
\label{elambda}
\ee

		Formulas
	(\ref{plambda})
		and
	(\ref{elambda})
                allow to use the quasiparticle interpretation
                for the spectrum of observables on Bethe
                vectors.  Each quasiparticle is created by
                operator
        $ B(\lam) $,
		it diminishes the
	$ S^3 $
		eigenvalue by
	$ 1 $
		and has momentum
        $ \pp(\lam) $
		and energy
        $ \epsilon(\lam) $
		given in
	(\ref{plambda})
		and
	(\ref{elambda}).
		Let us note, that
\be
        \epsilon(\lam) \, = \, \frac{1}{2} \,
                \frac{d}{d\lam} \pp(\lam) \;\;\; .
\ee
		The variable
        $ \lam $
                in this interpretation can be called a
                rapidity of a quasiparticle.

                It is possible to exclude the rapidity to get
                the dispersion relation, describing connection
                of energy and momentum
\be
	\epsilon(\pp) \, = \,
		\cos\pp \, - \, 1 \;\;\; .
\ee

                The eigenvalues of hamiltonian are all
                negative, so that the reference state
	$ \Omega $
                cannot be taken as a ground state, i.e. state
                of the lowest energy. It trivially changes if
                we take
	$ -\HAM $
		as a hamiltonian. Both cases
	$ \HAM $ and $ -\HAM $
                are interesting for the physical applications,
                corresponding to antiferromagnetic and
                ferromagnetic phases, correspondingly. The
                mathematical (and physical) features of the
	$ N \, \to \, \infty $
                limit in these two cases are completely
                different, as we shall see soon.

\section{	XXX${}_{1/2}$
                model.
                Physical spectrum in the ferromagnetic
                thermodynamic limit                     }
\label{sfive}

                In our case the thermodynamic limit is just
                limit
	$ N \rightarrow \infty $.
                We shall see, how BAE simplify in this limit.
                Looking at BAE
        (\ref{BAE})
		we see that
	$ N $
                enters there only in the exponent in the LHS.
                For real
        $ \lambda_1 $, \dots, $ \lambda_l $
		both sides in BAE
                are functions with values on the circle and
                LHS is wildly oscillating when
	$ N $
		is large. Taking the
	$ \log $
		we get
\be
        N \pp ( \lambda_j ) = 2 \pi Q_j +
                \sum_{k=1}^{l} \vph ( \lambda_j - \lambda_k )
        \; ,
\label{Nplam}
\ee
		where the integers
	$ Q_j $, $ 0 \leqslant Q_j \leqslant N-1 $
		define the branch of the
	$ \log $
		and
        $ \vph(\lambda) $
		is a fixed branch of
        $ \ln \frac { \lambda + i }
                    { \lambda - i } $.
		For large
	$ N $
		and
	$ Q $
		and fixed
	$ l $
		the second term in the RHS of
        (\ref{Nplam})
                is negligible and we get the usual
                quasicontinuous expression for the momentum of
                a free particle on a chain
\be
        \pp_j = 2\pi \frac {Q_j}
 			    {N}
        \; .
\ee
		In the ferromagnetic case when the hamiltonian is
	$ - \HAM $
                the energy of this particle is given by
        $ \epsilon(\pp) = 1 - \cos \pp $.

                The correction (second term in RHS of
        (\ref{Nplam}))
                expresses the scattering of these particles.
                The comparison with the usual quantum
                mechanical treatment of a particle in a box
                shows that
        $ \vph \left(\lambda_i - \lambda_k \right) $
                plays the role of the phase shift of particles
                with rapidities
        $ \lambda_j $
		and
        $ \lambda_k $.
		Thus the function
\be
        S ( \lambda - \mu ) =
                \frac { \lambda - \mu + i }
                      { \lambda - \mu - i }
\ee
		is a corresponding
        $ S $--matrix
		element.

		Another analogy is with the classical inverse
		scattering method, where the combination
\be
	Z(\lam) \, = \, B(\lam) \, A^{-1}(\lam)
\ee
		was more important, than
	$ B(\lam) $.
		The factor
	$ S $
		enters the exchange algebra
\be
	Z(\lam) \, Z(\mu) \, = \, Z(\mu)\, Z(\lam)\,
                S(\lam-\mu) \; ,
\ee
		valid in the limit
	$ N \to \infty $
		because the second term in the RHS of
	(\ref{ExchAB})
		effectively vanishes. Operator
	$ Z(\lam) $
		can be interpreted as a creation operator
		of a normalized particle state.

                This argument by analogy should be justified
                by the adequate scattering theory applicable
                to our case. We do not have time to do it here
                and refer to the original papers of Babbit and
                Thomas.

                The BAE allow also for the complex solutions
                which in our situation correspond to bound
                states.  Let us see it in more detail. The
                first nontrivial case is
        $ l = 2 $.
		From two BAE
\be
        \left ( \frac { \lambda_1 + {i}/{2} }
                      { \lambda_1 - {i}/{2} }
        \right )^N  =
                \frac   { \lambda_1 - \lambda_2 + i }
                        { \lambda_1 - \lambda_2 - i }
        \; ,
\label{BAEone}
\ee
\be
        \left ( \frac { \lambda_2 + {i}/{2} }
                      { \lambda_2 - {i}/{2} }
        \right )^N  =
                \frac   { \lambda_2 - \lambda_1 + i }
                        { \lambda_2 - \lambda_1 - i }
\ee
		we see, that
\be
        \left ( \frac { \lambda_1 + {i}/{2} }
                      { \lambda_1 - {i}/{2} }
        \right )^N
        \left ( \frac { \lambda_2 + {i}/{2} }
                      { \lambda_2 - {i}/{2} }
        \right )^N  = 1
                \; ,
\ee
		so that
        $ \pp ( \lambda_1 ) + \pp ( \lambda_2) $
                is real.
                Further, for
        $ \Imm \lambda_1 \neq 0 $
		the LHS in
        (\ref{BAEone})
		grows (or decreases) exponentially when
	$ N \rightarrow \infty $
                and to compensate it in the RHS we must have,
                that
\be
        \Imm ( \lambda_1 - \lambda_2 ) =
                i \;\;\;   ( \hbox{or } - i )
                \; .
\ee
		As
        $ \lambda_1 $
       		and
        $ \lambda_2 $
                can be interchanged, we can say,
                that in the limit
	$ N \rightarrow \infty $
        $ \lambda_1 $
       		and
        $ \lambda_2 $
                acquire the form
\be
        \lambda_1 = \lambda_{1/2} + \frac{i}{2} \, , \;\;
        \lambda_2 = \lambda_{1/2} - \frac{i}{2} \;\; ,
\ee
		where
        $ \lambda_{1/2} $
		is real. In the thermodynamic limit
        $ \lambda_{1/2} $
		can become arbitrary.

		The momentum
        $ \pp_{1/2}(\lambda) $
		and energy
        $ \epsilon_{1/2}(\lambda) $
                for the corresponding Bethe vector are given
                by
\be
        e^{i\pp_{1/2}(\lambda)}
	=
        e^{i\pp_0(\lambda+{i}/{2})
                + i\pp_0(\lambda-{i}/{2})}
	=
        \frac   {\lambda + i/2 + i/2}
                {\lambda - i/2 + i/2}
	\cdot
        \frac   {\lambda + i/2 - i/2}
                {\lambda - i/2 - i/2}
	=
        \frac   {\lambda + i}
                {\lambda - i}
\label{Bvecmom}
\ee
		and
\be
        \epsilon_{1/2}(\lambda)
	=
        \frac{1}{2}\frac{d}{d\lam}\ln\pp_{1/2}(\lambda)
	=
	\frac	{1}
                {\lambda^2 + 1}
               \; .
\label{Bvecen}
\ee
		The origin of notation
        $ \pp_0 (\lambda) $
		for the momentum
        (\ref{plambda})
		and
        $ \pp_{1/2}(\lambda) $, $ \epsilon_{1/2}(\lambda) $
                for momentum and energy of
		the complex solution will be clear soon.
                Excluding
        $ \lambda $
		from
        (\ref{Bvecmom})
		and
        (\ref{Bvecen})
		we get
\be
	\epsilon_{1/2} (\pp)
	=
	\frac{1}{2} ( 1 - \cos \pp )
        \; .
\ee
                The interpretation of this eigenvector as
                bound state is supported by the inequality
\be
        \epsilon_{1/2}(\pp) <
        \epsilon_0(\pp - \pp_1) + \epsilon_0(\pp_1)
\ee
		for all
        $ \pp $, $ \pp_1 $,
        $ \;\; 0 \leqslant \pp,\,\pp_1 \leqslant 2\pi $.
		The
        $ S $--matrix
                elements for scattering will be discussed
                later.

		For
	$ l > 2 $
                the complex solutions are described
                analogously.
                Roots
        $ \lambda_l $
		are combined in the complexes of type
	$ M $,
		where
	$ M $
                runs through half--integer values
	$ M = 0,1/2,1,\ldots \;\; $,
		defining the partition
\be
        l = \sum_{M} \nu_M ( 2M + 1)
        \; ,
\ee
		where
	$ \nu_M $
		gives the number of complexes of type
        $ M $.

		The set of integers
	$ \{ \nu_M \} $
                defines a configuration of Bethe roots.
                Each complex contains roots of the type
\be
        \lambda_{M,m} = \lambda_{M} + im \; , \;\;\;
		-M \leqslant m \leqslant M
                \; ,
\ee
		where
	$ \lam_M $ is real,
	$ m $
                being integer or half--integer together with
        $ M $.
		The momentum and energy of complex is given by
\be
        \pp_M(\lambda) = \frac{1}{i}
	\ln
                \frac   {\lambda + i(M+1/2)}
                        {\lambda - i(M+1/2)}
\label{compmom}
\ee
		and
\be
        \epsilon_M(\lambda)
	=
	\frac{1}{2}
	\frac	{2M+1}
                {\lambda^2 + (M+1/2)^2}
	=
	\frac	{1}
		{2M+1}
	(1 - \cos \pp_M)
        \; .
\label{dispM}
\ee
		The
        $ S $--matrix
		element for the scattering of complex of type
	$ 0 $
		on a complex of type
	$ M $
		is given by
\be
        S_{0,M}(\lambda)
	=
        \frac   {\lambda+iM}
                {\lambda-iM}   \cdot
        \frac   {\lambda+i(M+1)}
                {\lambda-i(M+1)}
\ee
		and for scattering of complexes
	$ M $
		and
	$ N $
\be
        S_{M,N}(\lambda)
	=
        \prod_{L=|M-N|}^{M+N} S_{0,L}(\lambda)
        \; .
\label{SMNlam}
\ee
                It is the superficial analogy of this formula
                with the Klebsch--Gordan formula for
        $ \sltwo $
		which prompted me to use label
	$ M $
                for complexes instead of their length
        $ 2M+1 $.

                The derivations are just the direct
                calculation of products
$$
         \prod_{m=-M}^{M}
                \frac   {\lambda+i/2+im}
                        {\lambda-i/2+im} \, , \;\;
         \prod_{m=-M}^{M}
                \frac   {\lambda+i+im}
                        {\lambda-i+im} \, , \;\;
         \prod_{m,n=-M,N}^{M,N}
                \frac   {\lambda+i+i(m+n)}
                        {\lambda-i+i(m+n)} \, ,
$$
                where many terms cancel.

                This finishes the description of the physical
                spectrum of
	$ -\HAM $
                in the thermodynamic limit.
                The ground state
	$ \Omega = \prod \omega_n $
                defines the incomplete infinite tensor product
                in the sense of John von Neumann.
                The space
        $ \HH_{\rm F} $
                is a completion of the states which
                differ from
	$ \omega_n $
		only in finite number of factors in
        $ \prod \otimes \hh_n $.
                The excitations are particles,
                classified by half--integers
	$ M $,
        $ M = 0,1/2,1,\ldots $
		and rapidity
        $ \lambda $
		(or momentum
        $ \pp $).
		The dispersion law for a particle of type
	$ M $
		is given by
        (\ref{dispM})
		and scattering matrix elements are given by
        (\ref{SMNlam}).
                The interpretation of particles with
	$ M > 0 $
		as bound states is possible but not necessary.

		The spin components
        $ S^{\pm} $
                have no sense in the physical Hilbert space
        $ \HH_{\rm F} $
		as they change vector
	$ \Omega $
		in any site
        $ n $.
		Operator
	$ S^3 $
		after shift by
	$ N/2 $
\be
        Q       =       \frac{N}{2} - S^{3}
\ee
		makes sense in
        $ \HH_{\rm F} $
		and has integer eigenvalues
        $ 2M+1 $.
                This phenomenon gives the example of
                ``symmetry breaking'' by
                vacuum: only
        $ U(1) $
                part of
        $ \sltwo $
                remains intact in the thermodynamic limit.
                Physically we call this phase ferromagnetic
                because the prescribed direction of spin
                (in 3-d direction in spin space) is macroscopic
                and fixed.

\section{XXX${}_{1/2}$ model. BAE for an arbitrary configuration}
\label{ssix}

		Before turning to physics of the
                antiferromagnetic chain we shall consider
		in more detail the BAE for arbitrary
		configuration
	$\{\nu_M\}$,
		where each integer
	$ \nu_M $
		gives the number of complexes of type
	$ M $
		combined from
	$ l $
		quasiparticles, so that
\be
	l=\sum_M(2M+1)\nu_M \;\;\; .
\ee
		We shall investigate the approximate BAE for
                the real centers of complexes
	$\lam_{M,i}$,
		and allow
	$l$
		to be of order
	$N/2$.
		There are some doubts,
                expressed in the literature, about validity
		of the picture of complexes in this
		case. Indeed, in our arguing above we supposed,
		that
	$l$
		is much smaller than
	$N$.
		However in the physical applications only
	$\nu_0$
		will be large
		and in this case the picture of complexes is correct.

		The BAE for the
	$\lam_{M,j}$, $j=1,\ldots,\nu_M$
		are obtained by multiplying
		the BAE for each complex in the LHS of
        (\ref{BAE})
                and rearranging the RHS
                according to the picture of complexes.
		They look as follows
\be
        e^{ip_M(\lam_{M,j})N}=\prod_{M'}\prod_{(M',k)\neq(M,j)}
                S_{M,M'}
		(\lam_{M,j}-\lam_{M',k}).
\label{BAEM}
\ee
		The factors in the RHS are scattering matrix elements from
        (\ref{SMNlam});
		in the LHS the momentum
	$\pp_M(\lam)$
		from
        (\ref{compmom})
		enters; condition
        $(M',k)\neq(M,j)$
		means, that among
	$\nu_{M'}$
		roots
	$\lam_{M',k}$,
		entering the RHS in
        (\ref{BAEM}),
		the one which is equal to
        $\lam_{M,j}$
                from LHS is absent.

                Taking the logarithm of
        (\ref{BAEM})
		and using the basic branch in the form
\be
         \frac{1}{i} \ln \frac{\lam+ia}{\lam-ia} =
                \pi - 2\arctg\frac{\lam}{a}
\ee
		we get the equation
\be
	2N\arctg\frac{\lam_{M,j}}{M+1/2} =
                2\pi Q_{M,j} +
                        \sum_{M'}\sum_{(M',k)\neq(M,j)}
                \Phi_{M,M'}(\lam_{M,j}-\lam_{M',k}) \; ,
\label{Narctg}
\ee
		where
\be
	\Phi_{M,M'}(\lam)=2\sum^{M+M'}_{L=|M-M'|}
		\left(\arctg\frac{\lam}{L}+\arctg\frac{\lam}
		{L+1}\right)
\ee
		with the understanding that the term with
	$L=0$
		is omitted and
	$Q_{M,j}$
                is an integer or half--integer (depending on
		the configuration,  which parametrizes
                the roots
	$\lam_{M,j}$).

		The main hypothesis of our investigation
		is that
	$Q_{M,j}$
		classify the roots uniquely and monotonously:
		roots
	$\lam_{M,j}$
		increase, when
	$Q_{M,j}$
                increase; moreover there are no coinciding
	$Q_{M,j}$
		for a given complex of type
	$M$.
		The last condition corresponds to the
		requirement for the BAE roots to be distinct,
		which was discussed in the course of
		derivation of BAE.

		We shall look for the real and bounded
		solutions of equation
        (\ref{Narctg}).
		For those the numbers
	$Q_{M,j}$
		have a natural bound. Indeed, taking into
		account, that
\be
	\arctg\pm\infty=\pm\frac{\pi}{2}
\ee
		and putting
	$\lam_{M,j}=\infty$
		we get for the corresponding
	$Q_{M,j}$
		the expression
\be
        Q_{M,\infty}=-\sum_{M'\neq M}
		\left(2\min(M,M')+1\right)\nu_{M'}-
               \left(2M+\frac12\right)(\nu_M-1)+
                        \frac{N}{2} \, .
\label{QMinf}
\ee
		The maximal admissible
	$Q_{M,j}$
		is then
\be
                Q_{M,\max}=Q_{M,\infty}-(2M+1)
\label{QMmax}
\ee
		because complex of type
	$M$
		has
	$(2M+1)$
                roots. We suppose, that when
	$Q_{M,j}$
		gets values bigger than
	$Q_{M,\max}$,
		the roots in our complex
		turn to be infinite one after another,
		so that for
	$Q_{M,j}=Q_{M,\infty}$
		the whole complex becomes infinite.

		I understand, that all this is quite
                a host of hypotheses, but the result
		we shall get soon is quite satisfactory.
		It will be nice to produce more
                detailed justification for our considerations.

		From
        (\ref{QMinf})
		and
        (\ref{QMmax})
		we get
\be
        Q_{M,\max}=\frac N2-\sum_{M'}
                J(M,M')\nu_{M'}-\frac12 \;\; ,
\ee
		where
\be
	J(M,M')=\left\{
\begin{array}{ll}
	2\min(M,M')+1 \quad & M\neq M'\\
	2M+\frac12 \quad & M=M' \;\; .
\end{array}
		\right.
\ee
		Analogously we find
	$Q_{M,\min}$.
		Due to the fact, that
	$ \arctg \lam $
		is odd we have
\be
	Q_{M,\min}=-Q_{M,\max} \;\; .
\ee
		Thus for the number of vacances
	$P_M$
		for the numbers
	$Q_{M,j}$
		we have
\be
        P_M=2Q_{M,\max}+1=N-2\sum_{M'} J(M,M')\nu_{M'} \;\; .
\ee
		The numbers
	$Q_{M,j}$
		are integers for odd
	$P_M$
                and half--integers for
	$P_M$
		even.

		Now we can estimate the number of Bethe
		vectors, characterized by the
		admissible numbers
	$Q_{M,j}$.
		For a given configuration
	$\{\nu_M\}$
		the states are given by fixing the
		distribution of
	$Q$-s
                over the vacances;
		so the whole number
	$Z(N,\{\nu_M\})$
		of them is given by
\be
        Z(N,\{\nu_M\})=\prod_M C^{\nu_M}_{P_M} \;\; ,
\ee
		where
	$C^m_n$
		is a binomial coefficient
\be
		C^m_n=\frac{n!}{m!(n-m)!} \;\; .
\ee
		Let us consider the number of states for
		given
	$l$
		and number of complexes
\be
	q=\sum\nu_M
\ee
		inside each configuration
	$\{\nu_M\}$
\be
        Z(N,l,q)=\sum_{\mbox {\scriptsize{ $
\begin{array}{c}
                \sum(2M+1)\nu_M=l \\
                \sum\nu_M=q
\end{array} $} } }
		Z\left(N;\{\nu_M\}\right) \;\; .
\ee
		We shall calculate
	$Z(N,l,q)$
		by reduction via a partial summation. For
		that we shall begin by extracting the
		contribution of roots of type
	$ 0 $,
		or in other words by substituting the
		configuration
	$\{\nu_M\}$
		by
	$\{\nu'_M\}$,
		where
\be
	\nu'_M=\nu_{M+1/2},\quad M=0,\frac12,\ldots \;\; .
\ee
	First we observe, that
\be
	P_M(N,\{\nu_M\})=P_{M-1/2}
		\left( N-2q, \{\nu'_M\}\right) \;\; .
\label{PMNnu}
\ee
		Indeed, it is easy to see that
\be
        J(M,M')=J(M-\frac12, M'-\frac12)+1 \;\; ,
\ee
		so that for
        $M\geqslant1/2$
\ba
        && P_M(N,\{\nu_M\})=N-2J(M,0)\nu_0-
        2\sum_{M'\geqslant1/2}J(M,M')\nu_{M'}=
\nonumber \\
       &&=N-2\nu_0-2\sum_{M'\geqslant 1/2} \left( J\left(M-\frac12, \
        M'-\frac12 \right)+1 \right)\nu_{M'}=
\nonumber \\
       &&=N-2q-2\sum_{M'}J \left( M-\frac12, M' \right) \nu'_{M'}
\ea
		and
        (\ref{PMNnu})
		follows. Thus we have a recurrence relation
\be
	Z(N,\{\nu_M\})=C^{\nu_0}_{P_0} Z
		\left( N-2q, \, \{\nu'_M\} \right)
\ee
		and summing over the allowed
	$\nu_0$
		we get
\be
	Z(N,l,q)=\sum^{q-1}_{\nu=0}
		C^{\nu}_{N-2q+\nu}Z(N-2q, \, l-q, \, q-\nu) \; .
\ee
		With the initial condition
\be
	Z(N,1,1)=N-1
\ee
		this gives
\be
	Z(N,l,q)=\frac{N-2l+1}{N-l+1}C^q_{N-l+1}C^q_{l-1}
\ee
		and finally for the number of the Bethe
		vectors with given
	$l$
\be
	Z(N,l)=\sum^l_{q=1}Z(N,l,q)=C^l_N-C^{l-1}_N \;\;.
\ee

                Now we remember that each Bethe vector of
		spin
	$\frac{N}{2}-l$
		is a highest weight in the multiplet of
		dimension
	$N-2l+1$.
		Thus the full number of
		states, described in our picture
\be
        Z=\sum_{l}(N-2l+1)Z(N,l)=2^N
\ee
		is equal to the dimension of our Hilbert
		space. This is very satisfactory
		and strongly confirms all the hypotheses,
		which we used in this
		calculation. We stop here the general
		investigation of the BAE
       (\ref{Narctg}).

\section{XXX${}_{1/2}$ model. Physical spectrum in the
		antiferromagnetic case}
\label{sseven}

                I am ready now to describe some important
                states. The ground state,
                i.e. the state of the lowest energy, is
		obtained by taking the maximal
                number of real roots. We shall suppose, that
        $N$
                is even to be able to have an
        $ \sltwo $
                invariant state.
		The corresponding
		configuration looks like
\be
		\nu_0=\frac N2\, ; \;\; \nu_M=0 \, ,
                        \; \;  M\geqslant\frac12.
\ee
		For this configuration
        $l=N/2$
		and so
\be
                S^{3}=\frac N2-l=0 \;\; ;
\ee
                thus the spin of the state vanishes.
		Now the number of vacances
	$P_0$
\be
		P_0=N-2J(0,0)\nu_0=N-\frac N2=\frac N2
\ee
		is equal to the number of roots and so
		there is no freedom for the
		allocating the numbers
	$Q_{0,k}$;
		they span all interval
\be
        -\frac N4 + \frac12 \leqslant
                Q_{0,k} \leqslant \frac N4-\frac12
\label{interv}
\ee
                being integer (half--integer) for
        $ N/2 $
		odd (even).

		Thus the state in question is unique and
		gives us the singlet for
        $\sltwo$
		group of spin observables.

		Using physical terminology we can call this
		state the Dirac sea of
		quasiparticles. The mere existence of this
		state is due to the Fermi
                character of the quasiparticles spectrum,
                i.e. to the condition that all
        $ Q_{0,j} $
                are distinct.

		In what follows we consider states, for which
	$\nu_0$
		differs from its
		maximal value
	$N/2$
		by a finite amount
\be
		\nu_0=\frac N2-\kappa \;\; .
\ee
		We shall see, how a Fock-like space of
		excitations will emerge step by
		step with increasing of
	$\kappa$
		in the limit
	$ N\to\infty $.
		Thus it is
	$\kappa$,
		which will play the role of the ``grading''
		in our definition of the
		physical portion in the formal infinite
		tensor product
	$\prod\otimes \CC^2$.

		For a fixed
	$\kappa$
		all
        $\nu_M$, $M\geqslant 1/2$
		are bounded, when
	$N\to\infty$.
		Indeed, from inequality
        $l\leqslant N/2$
		it follows
\be
        \sum_{M\geqslant1/2}(2M+1)\nu_M=
                l-\nu_0\leqslant\frac N2-\nu_0=\kappa
\ee
		and
	$N$
		disappears from this estimate.
		Thus for a fixed
	$\kappa$
		the number of configurations is finite and we
		can consider them one after another.

		For
	$\kappa=1$
		only
        $\nu_M=0$, $M \geqslant 1/2$ are allowed. The
	$l$ for these states is
	$l= N/2-1$, thus the spin is 1. The number of vacances
\be
	P_0=N-2\cdot\frac12\left(\frac N2-1\right)=\frac N2+1
\ee
                exceeds the number of roots by two.
                Thus two admissible numbers
	$Q_{0,j}$
		are not to be used, which gives the two
                parameter degeneracy of the state. In
		physical jargon one speaks of the holes
		in the Dirac sea.

		For
	$\kappa=2$
                there are two possibilities:
        $\nu_M=0$, $ M\geqslant1/2$ and
        $ \nu_{1/2}=1$, $\nu_M=0$, $M\geqslant1$.
		Let us consider the latter in more detail.
                First,
	$l= N/2$
		for it, so that the spin vanishes.
                Second, for the corresponding
        $P_{0}$
		we have
\be
	P_0=N-2\left( \frac N2-2 \right) \cdot\frac12-
		2J\left( 0,\frac12 \right)=\frac N2
\ee
		and
\be
	P_{1/2}=N-2\left(\frac N2-2\right)J\left(\frac12,
		0\right)-2J\left(\frac12,\frac12\right)= 4-3=1 \; .
\ee
		We see, that the number of vacances for real
		roots once more exceeds their
		number by 2 and there is no freedom at all for
		the root of complex of type
	$1/2$.

		In the former case
\be
	P_0=\frac N2-2\left(\frac N2-2\right)\frac12=\frac N2+2
\ee
		exceeds the number of roots by 4 and the
		spin of the state is 2.

		For general
	$\kappa$
                we have a similar picture. First, for the
		configurations
\be
        \nu_0=\frac N2-\kappa; \;\;\; \nu_M=0, \;\; M\geqslant1/2
\ee
		we have
\be
                P_0=\frac N2+\kappa \quad ,
\ee
		so that there are
	$2\kappa$
		holes, characterized by the missing places in
		the choice of admissible
	$Q_{0,j}$.
		The spin of this state is equal to $\kappa$.
		Then there are states with a smaller spin,
		corresponding to a few nonzero
        $\nu_M$, $M\geqslant1/2$.
		More on this will be said later.

		We return to a more detailed characteristic
		of the states already
		described. For this more control over the
		roots of BAE is needed.
		Fortunately these equations are simplified
		drastically in the thermodynamic
		limit
	$N\to\infty$.
		The real roots become quasicontinuous and
		we can
		evaluate their distribution.

		We begin with the ground state.
		The roots are real and corresponding
	$Q_{0,j}$
		fill without holes all interval
        (\ref{interv}).
		We shall put for
	$N/2$
		odd (with an evident correction for
	$N/2$
		even)
\be
                Q_{0,j} \, = \, j \;\; ,
\ee
		so that the BAE take the form
\be
	\arctg 2\lam_j=\frac{\pi j}N+\frac1N
		\sum_k\arctg (\lam_j-\lam_k).
\label{BAEarc}
\ee
		The variable
\be
                x=\frac{j}{N}
\ee
		becomes continuous in the limit
	$N\to\infty$
		with values
        $-1/4 \leqslant x\leqslant 1/4$;
		the set of roots
	$\lam_j$
		turn into function
	$\lam(x)$.

		The equation
       (\ref{BAEarc})
                becomes
\be
	\arctg 2\lam(x)=\pi x+
		\int^{1/4}_{-1/4}\arctg
			\left(\lam(x)-\lam(y)\right)dy
\label{BAEint}
\ee
		and looks rather formidable.
		Fortunately it is not
	$\lam(x)$
		which is of
		prime concern to us. Indeed,
		we are interested in the eigenvalues of local
		observables, which take the form
                of the sums over roots (see e.g. (\ref{eigmom})).
		With our conventions we have
\be
	\sum_j \ham(\lam_j)= N\int^{1/4}_{-1/4}\ham(\lam(x))dx=
                N\int^\infty_{-\infty}
                \ham(\lam)\rho(\lam)d\lam \;\; ,
\ee
		where the change of variables
	$\lam: \; x\to\lam(x)$
		maps interval
        $-1/4 \leqslant x \leqslant 1/4$
		into whole real line
	$-\infty<\lam<\infty$
		due to the monotonicity of
	$\lam(x)$.
		The density
	$\rho(\lam)$
		is nothing but
\be
	\rho(\lam)=\frac{dx}{d\lam}=
		\left. \frac1{\lam'(x)}
		\right|_{x=\lam^{-1}(\lam)} \;\; .
\ee

		Differentiating
        (\ref{BAEint})
                we get for this density, which we denote by
	$\rho_0(\lam)$
                for our state,
		a linear integral equation
\be
	\frac 2{1+4\lam^2}=\pi\rho_0(\lam)+
		\int^\infty_{-\infty}\frac{\rho_0(\mu)}
                {1+(\lam-\mu)^2}d\mu  \;\; ,
\ee
		which can be easily solved by
		Fourier transform. We get
\be
	\rho_0(\lam)=\frac1{2\cosh\pi\lam} \;\; .
\ee
		The momentum and energy of the ground
		state are given by
\be
	P_0=N\int \pp_0(\lam)\rho_0(\lam)d\lam=0
\ee
		due to the fact, that the integrand is
                odd if we use for
        $ \pp_0(\lam) $
                slightly shifted expression
\be
        \pp_0(\lam) = -2\arctg2\lam
\ee
                and
\be
	E_0=N\int\epsilon_0(\lam)\rho_0(\lam)d\lam=-N\ln 2 \;\; .
\ee
                Note, that the sign of
        $ \epsilon_{M}(\lam) $
                in this section is opposite to that in
                section
        \ref{sfive}.
		Thus the energy of the ground state is
		proportional to the volume as always
                in the correct thermodynamic limit.
		Adding it to the hamiltonian
	$\HAM$
		will make it nonnegative.

		Now we turn to the configuration
        $\nu_0=N/2-1,\;\; \nu_M=0,\; M\geqslant1/2$.
		There are two holes and we can put
\be
        Q_{0,j}=j+\theta(j-j_1)+\theta(j-j_2) \;\; ,
\ee
		where
	$\theta$
		is a step function
\be
                \theta(j)=\left\{ \begin{array}{cc}
                1\quad & j\geqslant 0\\
		0\quad &j<0
		\end{array} \right.
\ee
		and
	$j_1$
		and
	$j_2$
		are integer, characterizing the holes.
		By the same
                trick as before we get for the distribution
        $\rho_{\rm t}(\lam)$
        (t for triplet)
		of real roots the linear integral equations
\be
        \frac2{1+4\lam^2}=\pi\rho_{\rm t}(\lam)+
                \int^\infty_{-\infty}\frac{\rho_{\rm t}(\mu)}
		{1+(\lam-\mu)^2}d\mu+
        \frac\pi N(\dl(\lam-\lam_1)+\dl(\lam-\lam_2)) \; ,
\label{linint}
\ee
		where
	$\lam_1$
		and
	$\lam_2$
		are images of
	$x_1=j_1/N$ and $x_2=j_2/N$
		in the map
	$x\to\lam(x)$
		defined by
        $\lam_{\rm t}(x)$
		(or
	$\lam_0(x)$, because $\lam_1$ and
		$\lam_2$
		enter in terms of order
	$1/N$).
		From
        (\ref{linint})
		we get
\be
        \rho_{\rm t}(\lam)=\rho_0(\lam)+\frac1N(\si(\lam-\lam_1)+
                \si(\lam-\lam_2)) \;\; ,
\ee
		where
	$\si(\lam)$
		solves the equation
\be
	\si(\lam)+\frac1\pi\int^\infty_{-\infty}
		\frac{\si(\mu)}{1+(\lam-\mu)^2}d\mu+
	\dl(\lam)=0 \;\; .
\ee
		Solving it we can evaluate the momentum
		and energy of the corresponding state
\ba
               & P &=k(\lam_1)+k(\lam_2) \;\; ;
\label{Pklam}
\\
               & E &=E_0+ \ham(\lam_1)+\ham(\lam_2) \;\; ,
\label{Ehlam}
\ea
		where
\be
	 k(\lam)=\arctg\sinh\pi\lam, \ \ \
            \epsilon(\lam)=\frac\pi{2\cosh\pi\lam} \;\;.
\label{klam}
\ee

		It is time to comment, that the constructed
		states once  more allow for the
		particle interpretation: we have
		described a family of two particle states
		with the energy and momentum of a particle given by
        (\ref{klam})
		and dispersion law is
\be
        \epsilon(k)=\frac\pi2\cos k, \ \ \ \
                -\pi/2\leqslant k\leqslant\pi/2 \;\; .
\label{dispe}
\ee

		Next example is
        $\nu_0=N/2-2$, $\nu_{1/2}=1$, $\nu_M=0$, $M\geqslant1$.
		For the density of real roots
        $\rho_{\rm s}(\lam)$ (s for singlet)
		we get the equation
\ba
         \frac2{1+4\lam^{2}} &=& \pi\rho_{\rm s}(\lam)+
                \int^\infty_{-\infty}\frac{\rho_{\rm s}(\mu)}
                {1+(\lam-\mu)^2}d\mu+
\nonumber \\
        & + & \frac1N\left(
                \dl(\lam-\lam_1)+\dl(\lam-\lam_2)+
		\Phi'_{0,1/2}(\lam-\lam_{1/2})
                \right) \; ,
\label{denseq}
\ea
		where
	$\lam_1$ and $\lam_2$
		stand for the holes and the last term
		in the RHS is a contribution of the
		complex of type
	$ 1/2 $.
		For
	$\lam_{1/2}$
		we have one more equation
\be
        \arctg\lam_{1/2} = \frac1N\sum_j
                \Phi_{1/2,0}(\lam_{1/2}-\lam_{0,j}) \;\; ,
\label{arctgF}
\ee
		because the corresponding number
	$Q_{1/2,j}$
                has just one admissible
		value equal to zero.

		From
        (\ref{denseq})
		we get
\be
        \rho_{\rm s}(\lam)=\rho_0(\lam)+\frac1N
		(\si(\lam-\lam_1)+\si(\lam-\lam_2)+
                \om(\lam-\lam_{1/2})) \;\; ,
\ee
		where
	$\si(\lam)$
                is as above and
	$\om$
		is a solution of equation
\be
        \pi\om(\lam)+\int^\infty_{-\infty}
		\frac{\om(\mu)}{1+(\lam-\mu)^2}d\mu+
                \Phi'_{0,1/2}(\lam) = 0 \;\; .
\ee
		To evaluate
	$\lam_{1/2}$
		let us rewrite
        (\ref{arctgF})
		in the form
\ba
     &&  \arctg\lam_{1/2}-\int^\infty_{-\infty}\Phi_{1/2,0}
                (\lam_{1/2}-\lam)\rho_0(\lam)d\lam=
         \\
     &&  = \frac1N\int^\infty_{-\infty}\Phi_{1/2,0}
		(\lam_{1/2}-\lam)[\si(\lam-\lam_1)+\si
                (\lam-\lam_2)+\om(\lam-\lam_{1/2})]d\lam \; ,
\nonumber
\ea
		where the limit
	$N\to\infty$
		is already taken into account by changing the
		sums over
	$\lam_{0,j}$
		by integral over density
        $\rho_{\rm s}(\lam)$.
		LHS here vanishes for any
	$\lam_{1/2}$
		and contribution of
	$\om$
		disappears due to oddness of
		integrand ; so we get the equation
\be
	\int^\infty_{-\infty}\Phi_{1/2,0}(\lam_{1/2}-\lam)
                (\si(\lam-\lam_1)+
                \si(\lam-\lam_2))d\lam=0 \;\; ,
\ee
		or
\be
	\arctg 2(\lam_{1/2}-\lam_1)+
		\arctg 2(\lam_{1/2}-\lam_2)=0
\ee
		with the solution
\be
		\lam_{1/2}=\frac{\lam_1+\lam_2}2 \;\; .
\ee

                We are ready now to evaluate the observables.
                The spin of our state is zero.
		For momentum and energy we get exactly the
		same expressions
        (\ref{Pklam})
		and
        (\ref{Ehlam})
		as in the previous example; the contribution
		of a complex of type
	$1/2$
		cancels exactly.

                The examples considered are all, which give a
		two-parameter family of states.
		Returning to the particle interpretation,
		we can say, that our particles have
		spin
	$1/2$.
                Indeed, we constructed the highest weights in
		triplet and singlet
		two-particle states with exactly the same
		momentum and energy content. Thus
		they are highest weights in
	$\CC^2\otimes\CC^2$
                representation of a spin
		observable. This is why I say, that the
		particles have spin
	$1/2$.

		This picture is recurrently confirmed in the
		description of the next excitation.
		The state
        $\nu_0=N/2-\kappa$, $\nu_M=0$, $M\geqslant1/2$
		defines a
	$2\kappa$
		particle state being the
		highest weight in the highest spin irreducible
		component in
	$\prod^{2\ka}\otimes\CC^2$.
		All other states for the same
	$\ka$
		are states of
		lower spin, entering into multiplets with the
		number of particles not exceeding
        $ 2\ka $.
		The contribution of complexes of type
	$M$
		into energy and  momentum always vanishes,
		so that
		the energy-momentum expressions depend only
		on the number of particles.

		All this allows to say, that the only excitation
		of our system is
		a particle with spin 1/2 and energy -- momentum
		relation
        (\ref{dispe}).
		Note, that the momentum
	$\kappa(\lam)$
		runs through the half of
                the usual Brillouin zone. There is one important
		restriction: the
		number of particles  is even. These particles are
		usually referred
		to as spin waves. For a long time it was stated
		in the physical
                literature, that spin waves of the
		antiferromagnetic chain of
		spin 1/2 magnets has spin 1. Indeed, spin wave
		being a hole in
                the singlet Dirac sea corresponds to a turn
		of one spin,
                amounting to spin
	$1/2+1/2=1$.

		However, our more precise analysis shows, that
		turn of a spin
                corresponds to 2 holes and two spin wave
		excitations. The
		momentum of this state runs through the whole
                Brillouin zone
        $-\pi\leqslant k \leqslant \pi$.

		Mathematically the Hilbert space
        $\HH_{{\rm AF}}$
		is a (half) Fock
		space
\be
        \HH_{{\rm AF}}^{\even}=
        \sum^\infty_{n=0}\int^{\pi/2}_{-\pi/2}d\ka_1
                \ldots \int^{\pi/2}_{-\pi/2}d\ka_{2n}
		\prod^{2n}\otimes \CC^2 \;.
\ee
                (Compare it with Professor's Miwa lectures, where
        $\HH_{{\rm AF}}$
		corresponds to
        $\Lambda_0\otimes\Lambda_0$, or
                $\Lambda_1\otimes\Lambda_1)$.

                Natural question is how to describe
                one particle (or odd number of particles)
                state. The
		answer is, that they enter the chain of odd length.
		The lowest
                energy state there has spin 1/2 and have one hole
                in the
		distribution of numbers
	$Q_{o,j}$.
		Thus this state becomes a
                one-particle state in the thermodynamic limit.
		Chain of odd
		length has no ground state but just 1 particle,
		3 particles etc.
		states. The corresponding Hilbert space is
\be
        \HH^{\odd}_{{\rm AF}}=\sum^\infty_{n=0}
                \int_{-\pi/2}^{\pi/2} d\ka_1\ldots
                \int_{-\pi/2}^{\pi/2} d\ka_{2n+1}
                        \prod^{2n+1} \otimes \CC^2
\ee
		(and corresponds to
        $\Lambda_0\otimes\Lambda_1$ or
        $\Lambda_1\otimes\Lambda_0$
		in Professor's Miwa lectures).

		We finish with some formalization of our result.
		The expression
                for the Bethe state
\be
	\Phi(\{\lam\})=B(\lam_1)\ldots B(\lam_l)\Om
\ee
		can be formally rewritten as
\be
        \Phi(\{\lam\})= \left\{
        \exp\sum^l_{\lam_j=1}\ln B(\lam_i)
                        \right\} \Om
\ee
                and now it is possible to go to the
                thermodynamic limit. For
		the ground state $\Phi_0$ we have
\be
        \Phi_0=\exp \left\{
        N\int\limits^\infty_{-\infty}\ln B(\lam)
                \rho_0(\lam)d\lam
                        \right\} \Om \;\; .
\ee
                Exponential dependence of a true ground state
		on the volume is a
		typical phenomenon in quantum field theory.
		Now the triplet
		excited state
	$\Phi(\lam_1,\lam_2)$
		can be written as
\be
	\Phi(\lam_1,\lam_2)=\tilde Z(\lam_1)\tilde Z(\lam_2)\Phi_0
\ee
		without any reference to
	$\Om$
		and dependence on the volume. Here
\be
        \tilde Z(\lam)=\exp \left\{
        \int \ln B(\lam)\si(\lam-\mu)d\mu
                                \right\}
\ee
		plays the role of a creation operator
		of one-particle state from
		the physical ground state.

                As was mentioned in the section
        \ref{sfive}
                it is
\be
        Z(\lam)=\tilde Z(\lam) A^{-1}(\lam)
\label{Zlam}
\ee
                which is a more natural object in the
		scattering problem. Of
		course $Z(\lam)$ is defined up to a
		constant normalization
		factor, which we can use to cancel the
	$N$-dependent factor
	$a_\infty(\lam)$
\be
        a_\infty(\lam)=(\lam+\frac i2)^N\exp \left\{
		N\int^\infty_{-\infty}\ln\frac{
                \lam-\mu-i}{\lam-\mu}\rho_0(\mu)d\mu \right\}
\ee
		entering the eigenvalue
\be
		A_N(\lam)\Phi_0=a_\infty(\lam)\Phi_0
\ee
		in the limit $N\to\infty$, when the
		contribution of
	$D_N(\lam)$
		to
        $\Lambda(\lam,\{\lam\})$ vanishes.

                Operators
	$Z(\lam)$
                satisfy the exchange relation
\be
        Z(\lam)Z(\mu)=Z(\mu)Z(\lam)S_{\rm t}(\lam-\mu),
\ee
		where the phase-factor
        $S_{\rm t}(\lam-\mu)$
		is given by
\be
        S_{\rm t}(\lam)=\exp\left\{ \int^\infty_{-\infty}
                \ln\frac{\mu+i} {\mu-i}\si(\mu-\lam)d\mu
                        \right\}
	=\frac1i\frac{\Ga(\frac{1+i\lam}2)\Ga(1-\frac{i\lam}2)}
                {\Ga(\frac{1-i\lam}2)\Ga (1+\frac{i\lam}2)}\; .
\label{Stlam}
\ee
		In course of derivation the contribution
		of the second term in the
		exchange relations
        (\ref{ExchAB})
		is negleqted, which can be justified
		in the limit
	$N\to\infty$.

		The factor
        $S_{\rm t}(\lam)$
                is to be interpreted as a triplet
                eigenvalue of the
	$S$-matrix
		for spin 1/2 particles, acting in
	$\CC^2\otimes \CC^2$
\be
        S^{1/2,1/2}(\lam-\mu)=S_{\rm t}(\lam-\mu)
                        \left(\frac{\lam-\mu}{\lam-\mu+i}
		\Id+
                \frac {i}{\lam-\mu+i}\PP\right) \;\; .
\label{Shh}
\ee
		The creation operators
        $Z_\veps(\lam)$, $\veps=\pm1$
		are to satisfy
		the exchange Zamolodchikov relation
\be
        Z_{\veps_1}(\lam_1)
            Z_{\veps_2}(\lam_2)=Z_{\veps'_2}(\lam_2)
      Z_{\veps'_1}(\lam_1)S^{\veps'_1\veps'_2}_{\veps_1 \veps_2}
                (\lam-\mu) \; .
\ee
		We did not construct operator
        $Z_\veps(\lam)$
                (the vertex operators
                of the second kind in Professor Miwa
                terminology). We can only
                identify in
        $\HH^{\even}_{{\rm AF}}$
\be
        Z_+(\lam)Z_+(\mu)=Z(\lam)Z(\mu)  \quad ,
\ee
		where
	$Z(\lam)$ is given by
        (\ref{Zlam})
		and
\ba
     && Z_+(\lam) Z_-(\mu)-Z_-(\lam)Z_+(\mu)=
         \\
          &&       Z(\lam)\exp \left\{
                \int^\infty_{-\infty}\ln B(\si)
                \omega \left( \frac{\lam+\mu}2-\si
                        \right) d\si \right\}
                B\left( \frac{\lam+\mu+i}2
                        \right) B\left(
                \frac{\lam+\mu-i}2 \right) Z(\mu) .
\nonumber
\ea
		This identification is, however, sufficient
		to justify all
	 $S$-matrix
        (\ref{Shh}).

		The interesting but not understood comment
		on the formula
        (\ref{Stlam})
		is as follows: the phase-factor
        $S_{\rm t}(\lam)$ coincides with the
	$S$-matrix for the rotationally symmetric
		subspace of the
                Laplacian on Poincare plane. Indeed putting
        $s=(1+i\lam)/2$
		we get
\be
                S_{\rm t}(\lam)=\frac{f(s)}{f(1-s)}  \quad ,
\ee
		where
\be
                f(s)=\frac{\Ga(s)}{\Ga(1/2+s)}
\ee
		is a Harrish--Chandra factor for
        $\sltwor$.
		With this
		intriguing comment we finish our long and
		detailed treatment of
		the
	XXX${}_{1/2}$
                model. From now on I shall describe several
                directions of development and/or generalization
		along the
		similar lines without giving too much details.
		The first
		generalization is a XXX model for higher spin.

\section{XXX${}_s$ model}
\label{seight}

		Now I consider the spin chain with local spin
		variables
	$S^\al_n$
		realizing the finite dimensional
		representation of
        $\sltwo$ in $2s+1$
		dimensional space
	$\CC^{2s+1}$,
		where
	$s$
                is spin, integer or half--integer.  I am not
		ready to write the
		corresponding hamiltonian. To maintain the
		integrability I shall
		find it as a member of the commuting family
		of operators,
		generating function for which will be trace
		of an appropriate
		monodromy of the family of local Lax
		operators, satisfying the
                FCR {\it a-l\'a}
        (\ref{RTT}).

		The Lax operator $L_{n,a}(\lam)$ with the
		auxilialy space
	$\VV=\CC^2$
		does not differ from
        (\ref{Laxlam}).
		Indeed, operator, defined in
	$\hh_n\otimes\VV=\CC^{2s+1}\otimes\CC^2$
		by matrix
\be
        L_{n,a}(\lam)=\lam \Id+i\sum_{\al}S^\al_n\si^{\al} =
		\left(\begin{array}{cc} \lam+iS^3_n&
		iS^-_n\\
		iS^+_n&\lam-iS^3_n
	\end{array} \right)
\label{Laxs}
\ee
		satisfy the relation
\be
        R_{a_1,a_2}(\lam-\mu)L_{n,a_1}(\lam) L_{n,a_2}(\mu)=
           L_{n,a_2}(\mu)L_{n,a_1}(\lam) R_{a_1,a_2}(\lam-\mu)
\label{FCRs}
\ee
		with the same
        $R$-matrix
        $R_{a_1,a_2}(\lam)$ from
        (\ref{Laxtwotwo}).
                The derivation from section
        \ref{sthree}
                is not applicable.
                We shall not derive
        (\ref{FCRs})
                here because a more general check will
		be done below
		for the XXZ model.

		Introducing the monodromy
\be
	T_a(\lam)=\prod L_{n,a}(\lam)=
		\left(\begin{array}{cc} A(\lam)&B(\lam)\\
		C(\lam)&D(\lam)
	\end{array} \right)
\ee
		we see, that it satisfies FCR of the form
        (\ref{RTT}),
		so that its trace
\be
		F(\lam)=A(\lam)+D(\lam)
\ee
		is a commuting family of operators
\be
		[F(\lam),F(\mu)]=0.
\ee
		This family can be diagonalized by
		means of Algebraic Bethe
		Ansatz (ABA). Indeed, we have local vacuum
	$\om_n$ -- the highest
		weight in
        $\CC^{2s+1}$, the reference state
\be
		\Om=\prod \otimes \om_n
\ee
		with the eigenvalues for
	$A(\lam)$, $D(\lam)$ and $C(\lam)$
\be
                A(\lam)\Om=\alpha^N(\lam)\Om,\quad
                D(\lam)\Om=\delta^N(\lam)\Om,\quad
                C(\lam)\Om=0 \;\; ,
\ee
		where
\be
                \alpha(\lam)=\lam+is \; ,
                        \quad \delta(\lam)=\lam-is \;\; ,
\ee
		and exactly the same exchange relation for
	$A(\lam)$, $D(\lam)$
		and $B(\lam)$ as
        (\ref{ExchAB})-(\ref{ExchDB}).
		Thus the state
\be
		\Phi(\{\lam\})=B(\lam_1)\ldots B(\lam_l)\Om
\ee
		is an eigenstate of the family
	$F(\lam)$
		with the eigenvalue
\be
        \Lambda(\lam,\{\lam\})=(\lam+is)^N
		\prod^l_{j=1}\frac{\lam-\lam_j-i}{\lam-\lam_j}+
		(\lam-is)^N
                \prod^l_{i=1}\frac{\lam-\lam_j+i}{\lam-\lam_j} \; ,
\ee
		if $\{\lam\}$ are roots of the BAE
\be
        \left(\frac{\lam_k+is}{\lam_k-is}\right)^N=
	  \prod^l_{j\neq k}
                \frac{\lam_k-\lam_j+i}{\lam_k-\lam_j-i} \quad .
\label{BAEs}
\ee
		When
	$s=1/2$
		we return to the case already considered above.

		However the construction of the local
                hamiltonian cannot repeat
                one from section
        \ref{sthree}.
                Indeed, in no point
	$\lam$
		the Lax operator
	$L_{n,a}(\lam)$
                is a permutation, just because the quantum space
	$\hh_n=\CC^{2s+1}$
		and auxiliary space
	$\VV=\CC^2$ are
		essentially different.

		The way out is to find another
                Lax operator, for which the auxiliary space
        $\VV$ is $\CC^{2s+1}$.

		The existence of such an operator is
		based on a more general
		interpretation of the FCR due to V.Drinfeld.

		In this interpretation the generating
		object for Lax operators is a
		universal
        $R$-matrix
        $\RMA$
		defined as an element in
        $ \Acr \otimes \Acr $
		for some algebra
        $ \Acr $,
		satisfying the abstract Yang--Baxter
		relation (YBR)
\be
                \RMA_{12}\RMA_{13}\RMA_{23}=
                \RMA_{23}\RMA_{13}\RMA_{12} \;\; .
\label{YBR}
\ee

		This equation holds in
        $\Acr \otimes \Acr \otimes \Acr$
		and rather evident
		notations are used, i.e.
\be
        \RMA_{12}=\RMA\otimes \Id,\quad
                \RMA_{23} =\Id\otimes \RMA  \quad .
\ee
		The algebra
        $\Acr$ must have a family of representations
	$\rho(\lam,a)$,
                parametrized by a discrete label
        $a$ and
		continuous parameter
	$\lam$.
                For instance the loop algebra of any
		unitary group has such representations
		called the evaluation
		representations. In our case of XXX models
		the algebra
        $\Acr$
                was identified by Drinfeld and called
		Yangian by him. Below on
		the case of XXZ models we will encounter the
	$q$-deformed
		affine algebra as
        $\Acr$.

		The concrete Lax operators are obtained
		via the evaluation
		representations of the universal
	$\RMA$-matrix, i.e.
\be
        L_{n,a}(\lam-\mu)=(\rho(a,\lam)\otimes\rho(n,\mu))\RMA
                = R^{a,n}(\lam-\mu)    \quad .
\ee
		The dependence in the LHS on
	$\lam-\mu$
		reflects some
		homogeneity in the family of representations
	$\rho(a,\lam)$.
                The Yangian for
        $\sltwo$
		has representations
	$\rho(a,\lam)$,
		where
	$a$
		is just spin label of representations of
        $\sltwo$,
        $a=0,1/2,1,\ldots \;$.
		The relation
        (\ref{FCR})
		is obtained if we apply the representation
	$\rho(1/2,\lam)\otimes
        \rho(1/2,\mu) \otimes \rho(s,\si)$
		to YBR
        (\ref{YBR}),
		put
	$\si=0$
                and identify
\be
        R_{a_1,a_2}(\lam) = R^{1/2,1/2}(\lam)\; ; \quad
          L_{n,a_1}(\lam) = R^{1/2,s}(\lam).
\ee
		However we can use another combination
		of the representations.
		Let us rewrite the YBR
        (\ref{YBR})
		in the form
\be
                \RMA_{12}\RMA_{32}\RMA_{31}=
                \RMA_{31}\RMA_{32}\RMA_{12} \;\; ,
\label{YBRtoo}
\ee
		which can be easily derived from
        (\ref{YBR})
		by applying the appropriate permutation to
                it.
		Now apply to
        (\ref{YBRtoo})
		the representation
	$\rho(s_1,\lam)\otimes \rho(s_2,\mu)\otimes
		\rho(1/2,\si)$.
		We get
\ba
        R^{s_1,s_2}(\lam-\mu)R^{1/2,s_2}(\si-\mu)R^{1/2,s_1}(\si-\lam)=
\nonumber \\
	=R^{1/2,s_1}(\si-\lam)R^{1/2,s_2}(\si-\mu) R^{s_1,s_2}(\lam-\mu)
	  \;\; .
\label{YBRhalf}
\ea
		The factors
	$R^{1/2,s}(\lam)$
		can be identified with the Lax
		operators
	$L_{n,a}(\lam)$
		above; the operator
	$R^{s_1,s_2}(\lam)$
		give us a new Lax operator; we can take
		representation with spin
	$s_1$
		as a local quantum space and that with spin
	$s_2$
		as an auxiliary space. In particular for
	$s_1=s_2$
		we get the Lax
		operator
	$L_{n,f}(\lam)$
		we are looking for. Equation
       (\ref{YBRhalf})
		is a linear equation, which allows to
		calculate
	$L_{n,f}$ if
	$L_{n,a}$
		are known. We shall call
	$L_{n,f}$
		the fundamental Lax
		operator and label
	$f$
		stems from this.

		Let us now calculate
	$R^{s_1,s_2}(\lam)$
		for representations
	$s_1$ and $s_2$
		being the same, using equation
        (\ref{YBRhalf}).
		To simplify the notation we shall denote
		two sets of spin variables by
	$S^\al$ and $T^\al$
		and use the notations
\be
	L_T(\lam)=\lam+i(T,\si),\quad
		L_S(\lam)=\lam+i(S,\si)
\ee
		for the corresponding Lax operators
	$R^{1/2,s}$.
		Here
\be
        (T,\si)=\sum_{\al} T^\al \si^\al
\ee
		and analogously for
	$(S,\si)$.
		We shall look for
	$R^{s_1,s_2}(\lam)$
		in the form
\be
	R^{s_1,s_2}(\lam)=\PP^{s_1,s_2}
                r(( S, T),\lam) \;\; ,
\ee
		where
	$\PP^{s_1,s_2}$
		is a permutation in
	$\CC^{2s+1} \otimes \CC^{2s+1}$ and
        $( S, T)$
                is a Casimir
        $ C $
\be
        C = ( S, T)= \sum_\al S^\al T^\al \;\; .
\ee
		Using
\be
	\PP(S,\si)\PP=(T,\si)
\ee
		we rewrite the equation
        (\ref{YBRhalf})
		as follows
\be
        (\lam-i(T,\si))(\mu-i(S,\si)) r(\lam-\mu)=
                r(\lam-\mu)(\mu-i(T,\si))(\lam-i(S,\si)) \, .
\label{TSr}
\ee
		We have due to the property of Pauli matrices
	$\si^\al$
\be
        (T,\si)(S,\si)=(T,S)+i(( S\times T),\si) \quad ,
\ee
		where
\be
        ( S\times T)^{\al}=\veps_{\al\bet\ga}
                S^{\bet} T^{\ga}
		\;\; .
\ee
		Now using the central property of Casimir
		we transform equation
        (\ref{TSr})
		into
\be
        (\lam S^{\al}+(T\times S)^{\al})r(\lam)=
                r(\lam)(\lam T^{\al}+(T\times S)^{\al}) \; .
\label{Srlam}
\ee
		Due to symmetry, it is enough to consider
		one out of three equations
        (\ref{Srlam})
                e.g. the combination
\be
        (\lam S^{+}+i(T^{3}S^{+}-S^{3}T^{+})) r(\lam)=
                r(\lam)(\lam T^{+}+i(T^{3}S^{+}-S^{3}T^{+})) \;.
\label{Splusr}
\ee
		We shall use a convenient variable
	$J$
		instead of Casimir
	$(S, T)$;
		taking into account that representations for
	$S$ and $T$
		are irreducible we have
\be
	(S+T)^2 = S^2+T^2+ 2(S,T)=
                2s(s+1) + 2(S,T)=J(J+1) \; ,
\ee
		where the operator
	$ J $
		have an eigenvalue
	$j$
		in each irreducible representation
	$D_j$
		entering the Klebsch--Gordan decomposition
\be
        D_s\otimes D_s = \sum^{2s}_{j=0} D_j \;\; .
\ee
		We shall look for operator
	$r(\lam)$
		as a function of
	$J$.
		To find it we shall use the equation
        (\ref{Splusr})
                in the subspace of the highest weights in
                each
        $ D_{j} $,
                i.e. put
\be
                T^{+} + S^{+} = 0 \;\; .
\ee
		This is permissible because
\be
        [T^{+} S^{3} - S^{+} T^{3} \; ,\; T^{+} + S^{+}]=0 \;\; .
\ee
		In this subspace
		due to the fact, that in general
\be
        (S+T)^2 = (S^3+T^3)^2+S^3+T^3 + (S^-+T^-)(S^++T^+)
\ee
                we can identify
\be
                J=S^3+T^3      \quad .
\ee
                In the constrainted subspace the equation
        (\ref{Splusr})
                reduces to
\be
        (\lam S^+ + i JS^+)r(\lam,J)=r(\lam,J)(-\lam S^+ + i JS^+)
\label{JSr}
\ee
		Now we use the commutation relation
\be
        S^+J=S^+(S^3 + T^3) = (S^3+T^3-1)S^+=(J-1)S^+
\ee
		to turn
        (\ref{JSr})
		into the functional equation
\be
	(\lam+i J)r(\lam,J-1)=r(\lam,J)(-\lam+i J)
\label{functeq}
\ee
		with solution
\be
        r(J,\lam) =
                \frac{\Ga(J+1+i\lam)}{\Ga(J+1-i\lam)}\;\; ,
\label{rJlam}
\ee
		normalized in such a way
\be
	r(J,0)=I; \quad r(J,-\lam)r(J,\lam)=\Id \;\;.
\ee
		Of course in our case of finite dimensional
        $D_s$
                we are interested only in
	$r(J,\lam)$ for $J$
		taking values
        $J=0,1,2,\ldots, 2s$.
		But in what follows we shall use the
		representations with any complex
	$s$
		and then formula
        (\ref{rJlam})
		will be used in all generality.

		Having this Lax operator
	$L_{n,f}(\lam)$
		we can write one more variant of FCR
\be
	R_{f_1,f_2}(\lam-\mu)L_{n,f_1}(\lam) L_{n,f_2}(\mu)=
            L_{n,f_2}(\mu)L_{n,f_1}(\lam)
                R_{f_1,f_2}(\lam-\mu) \; .
\ee
		From this we infer, that the spectral
		invariants of the monodromy
\be
	T_f(\lam)=L_{N,f}(\lam)L_{N-1,f}(\lam)\ldots L_{1,f}(\lam)
\ee
		(i.e.
        $F_f(\lam)=\tr_{f}T_f(\lam)$)
		are commuting
\be
                [F_f(\lam),F_f(\mu)]=0 \;\; .
\ee
		The relation
        (\ref{YBRhalf}) with
        $ s_{1} = s_{2} $,
		where
        $R^{1/2,s}(\lam-\mu)$
		is used as
        $R$-matrix and
        $R^{1/2,s}(\mu)$
		and
        $R^{s,s}(\lam)$
		as Lax operators leads to the commutativity of
		the families
	$F_f(\lam)$ and $F_a(\lam)$
\be
		[F_f(\lam),F_a(\mu)]=0 \;\; .
\ee
		Thus we can use
	$F_f(\lam)$
                to get observables and
	$F_a(\lam)$
                to construct BAE.

                Repeating the considerations in section
        \ref{sthree}
		we get
\be
		F_f(0)=U=e^{i\PMO}
\ee
		and
\be
        H=i\left.\frac d{d\lam}\ln F_f(\lam)\right|_{\lam=0}
                =\sum^N_{n=1}\HAM_{n,n+1}   \quad ,
\ee
		where
\be
        \HAM_{n,n+1}=i \left.\frac d{d\lam}\ln r(J,\lam)
                        \right|_{\lam=0} \quad ,
\label{Hamn}
\ee
		where
	$J$
		is constructed via local spins
	$S^\al_n$ and
	$S^\al_{n+1}$ as
\be
	J(J+1)=2\sum_\al (S^\al_n S^\al_{n+1})+2s(s+1)
\ee
		From
        (\ref{rJlam})
		and
        (\ref{Hamn})
		we get
\be
        \HAM_{n,n+1}=-2\psi(J+1)  \quad ,
\ee
		where
	$\psi(z)$
		is logarithmic derivative of $\Ga(z)$.
		For positive integer
	$n$
		we have
\be
        \psi(1+n)=\sum_{k=1}^n \frac1k-\ga \;\; ,
\ee
		where
	$\ga$
                is the Euler constant, and it allows to express
	$\HAM_{n,n+1}$
		as a polinomial in invariant
	$\sum_\al
		S^\al_n S^\al_{n+1}=C_{n,n+1}$ as follows
\be
	\HAM_{n,n+1}=\sum^{2s}_{j=1}c_k(C_{n,n+1})^k
                =f_{2s}(C_{n,n+1}) \;\; ,
\label{HamnC}
\ee
		where the polinomial
	$f_{2s}(x)$
                can be written using Lagrange interpolation
                as
\be
        f_{2s}(x)=\sum^{2s}_{j=1}\left(\sum^j_{k=1}\frac1k
                        - \gamma \right)
                \prod^{2s}_{l=0}\frac{x-x_l}{x_j-x_l}, \;\;\;
                x_l=\frac12(l(l+1)-2s(s+1)) \; .
\label{fsx}
\ee
		In particular for
	$s=1$
		we have
\be
                c_1 \, = \, - c_2 \;\; ,
\ee
                so that the hamiltonian
\be
	 \HAM=\sum_{\al,n}\left(S^\al_{n}S^\al_{n+1}-
			(S^\al_n S^{\al}_{n+1})^2\right)
\ee
                is integrable for the representation of
		local spins in
	$\CC^3$.
		The naive generalization of the hamiltonian
        (\ref{Hamhalf})
		by simple substitution of operators of spins
		1/2 by those of spin
		1 is not integrable.

		The construction of the integrable hamiltonians
                for spin $ s $
                magnetic chains is one of real achievements of
                the Algebraic Bethe Ansatz. Indeed, without
		well understood connection of
		integrable hamiltonians and Lax operators there
		is no hope to reproduce the formulas
        (\ref{HamnC}),
        (\ref{fsx}).

                Now I shall give without derivation the
                expression for the eigenvalue
        $\Lambda_f(\lam,\{\lam\})$
		of family
	$F_f(\lam)$
		on the Bethe vector
		$\Phi(\{\lam\})$:
\be
        \Lambda_{f}(\lam,\{\lam\})=
                \sum^s_{m=-s} \alpha_m(\lam)^N
			\prod^l_{k=1}
                c_m(\lam-\lam_k)  \;\; .
\label{Lamlam}
\ee
                The complete list of
        $ \alpha_{m}(\lam) $
                and
        $ c_{m}(\lam) $
                is not important. Suffice to say, that
\be
        \alpha_m(0) \, = \, 0
\ee
                for all
        $ m = -s, \, -s+1, \ldots s-1 $ and
        $ \alpha_{s}(0)=1 $.
                Further,
\be
        c_s(\lam)=\frac{\lam-is}{\lam+is} \;\; .
\ee
                The derivation is based on the relation
        (\ref{YBRhalf})
		which allows to commute the diagonal
		elements of matrix
	$T_f(\lam)$
		with
	$B(\lam)$
		from
	$T_a(\lam)$.
		From
        (\ref{Lamlam})
		we read the momentum and energy of
		quasiparticles
\ba
        \pp(\lam) & = & \frac{1}{i} \ln
                \frac{\lam+is}{\lam-is} \quad ;  \\
        \ep(\lam) & = & - \frac{s}{\lam^{2}+s^{2}} \;\; .
\ea
		The relation
\be
        \ep(\lam)= \frac{1}{2} \frac d{d\lam} \pp(\lam)
\ee
		holds; also it is
	$\pp(\lam)$
		which enters the LHS of BAE
        (\ref{BAEs}),
		so we do not need the Lax operator
        $L_{n,f}(\lam)$
		to calculate
	$\pp(\lam)$ and $\ep(\lam)$.

		The BAE
        (\ref{BAEs})
		can be investigated similarly to what was
		done in case
	$s=1/2$.
		In the limit
	$N\to\infty$
		the roots are combined into complexes of type
	$M$.
                The momenta of these complexes are
		distinct of those for spin 1/2. The
	$S$-matrices
		are defined by the RHS of
		BAE and are the same for any spin.
		One can make the counting of
		roots and get completeness by showing,
		that full number of
		states (for which Bethe vectors are the
		highest weights) is
		equal to
	$(2s+1)^N$.

		Finally, we add some comments on the
                thermodynamic limit in the
		antiferromagnetic case. The ground
                state
        $ \Phi_{0} $
                is given by
        $\nu_{s-1/2}= N/2 $,
	$\nu_M=0$,
	$M\neq s-1/2$.
		The excitations
		correspond to
	$\nu_{s-1/2}$
		macroscopic and all other
	$\nu_M$
		finite. They have particle interpretation
		as spin 1/2 particles
		with the same one particle momentum and
		energy as in the case
	$s=1/2$.
		However the counting of their states shows,
                that excitations have more degrees of freedom,
                than just rapidity
        $\lam$ and spin $\veps$.

		I shall describe the picture of excitations
		(without derivation)
		using the language of the creation operators.
		In addition to spin label
        $\veps$
		and rapidity
	$\lam$
		this operator is supplied by
                a pair of indeces
        $a,a'$
		assuming integer values from 0 to
	$2s$
		and subject to condition
	$a'=a\pm1$.
		The $n$-particle excitation
		is given by a ``string''
\be
        Z^{a_0,a_1}_{\veps_1}(\lam_1) Z^{a_1,a_2}_{\veps_2}(\lam_2)
		\ldots
        Z^{a_{n-1}, a_n}_{\veps_n}(\lam_n )\Phi_0   \;\; ,
\ee
		where
	$a_0=0$ and $a_n=0$.
		The counting of Bethe vectors is
                in exact accord with this picture.

                The exchange relation for operators
        $ Z $
                employs  the
	$S$-matrix,
		which is a tensor product of the spin 1/2
	$S$-matrix
                from section
        \ref{sseven}
                and
	$S$-matrix,
                which acts on the indeces $a$.

		The latter will be denoted by
        $S\left( \mbox{\scriptsize{$
\begin{array}{c|ccc}
                &&b&\\
           \lam& a&& d\\
                &&c&
\end{array} $}}
	\right)$
		and it enters the exchange relations
		as follows
\be
	Z^{ab}(\lam)Z^{bc}(\mu)=
		\sum_d	Z^{ad}(\mu)Z^{dc}(\lam)
        S\left( \mbox{\scriptsize{$
\begin{array}{c|ccc}
                    &&b&\\
           \lam-\mu& a&& c\\
                    &&d&
\end{array}     $}}
        \right) \;\; ,
\label{ZabZbc}
\ee
		where we suppressed the usual spin
		variables. They are easily
		introduced if we write the full
	$S$-matrix as
\be
		S=S^{1/2,1/2}(\lam)\otimes
        S\left( \mbox{\scriptsize{$
\begin{array}{c|ccc}
		&&b&\\
	   \lam& a&& d\\
		&&c&
\end{array}   $}}
        \right) \;\; .
\ee
		The consistency of relations of type
        (\ref{ZabZbc})
		is based on a star-tri\-an\-gle rela\-tion for
       $ S\left( \mbox{\scriptsize{$
\begin{array}{c|ccc}
		&&b&\\
	   \lam& a&& d\\
		&&c&
\end{array}   $}}
	\right) $
\ba
       && \sum_{p}
        S\left( \mbox{\scriptsize{$
\begin{array}{c|ccc}
                   &   & b &  \\
           \lam-\mu& a &   & c\\
                   &   & p &
\end{array}   $}}
        \right) \:
        S\left( \mbox{\scriptsize{$
\begin{array}{c|ccc}
                   &   & c &  \\
     \lam-\sigma   & p &   & d\\
                   &   & e &
\end{array} $}}
        \right) \:
        S\left( \mbox{\scriptsize{$
\begin{array}{c|ccc}
                   &   & p &  \\
     \mu -\sigma   & a &   & e\\
                   &   & f &
\end{array}  $}}
        \right) \; =
\nonumber \\    && = \;
           \sum_{p}
        S\left( \mbox{\scriptsize{$
\begin{array}{c|ccc}
                   &   & c &  \\
         \mu-\sigma& b &   & d\\
                   &   & p &
\end{array}  $}}
        \right) \:
        S\left( \mbox{\scriptsize{$
\begin{array}{c|ccc}
                   &   & b &  \\
     \lam-\sigma   & a &   & p\\
                   &   & f &
\end{array} $}}
        \right) \:
        S\left( \mbox{\scriptsize{$
\begin{array}{c|ccc}
                   &   & p &  \\
     \lam - \mu    & f &   & d\\
                   &   & e &
\end{array}  $}}
        \right) \: .
\nonumber
\ea
		The explicit expression for
       $ S\left( \mbox{\scriptsize{$
\begin{array}{c|ccc}
		&&b&\\
	   \lam& a&& d\\
		&&c&
\end{array}  $}}
	\right) $
                contains factor
        $S_{0}(\lam)$
                given by
\be
        S_{0}(\lam)=\exp\left\{-i\int^\infty_0\frac{dx}x
                \frac{\sin(\lam x) \sinh(s+1/2)x}
                {\cosh(x/2) \sinh(s+1)x}
			\right\} \;\; .
\label{Szlam}
\ee
                In particular
\be
        S \left( \mbox{\scriptsize{$
\begin{array}{c|ccc}
		&&a+1&\\
	   \lam& a&& a+2\\
		&&a+1&
\end{array} $}}
	\right)
         =
        S\left( \mbox{\scriptsize{$
\begin{array}{c|ccc}
		&&a-1&\\
	   \lam& a&& a-2\\
		&&a-1&
\end{array} $}}
	\right)
                = S_0(\lam) \; ;
\ee
                and
\be
        S\left( \mbox{\scriptsize{$
\begin{array}{c|ccc}
		&&a+1&\\
	   \lam& a&& a\\
		&&a+1&
\end{array}   $}}
	\right)
         =
        \frac{
              \sinh\left(\frac\pi{2s+2}(\lam+i(a+1))\right)
                        \sin\frac{\pi}{2s+2}}
             {
                \sin\left(\frac\pi{2s+2}(a+1)\right)
                \sinh\frac\pi{2s+2}(\lam-i)}
                        S_{0}(\lam)  \; .
\label{Slamb}
\ee
                Other components
\be
        S\left( \mbox{\scriptsize{$
\begin{array}{c|ccc}
		&&a-1&\\
	   \lam& a&& a\\
		&&a-1&
\end{array} $}}
        \right), \quad
        S\left( \mbox{\scriptsize{$
\begin{array}{c|ccc}
		&&a-1&\\
	   \lam& a&& a\\
		&&a+1&
\end{array} $}}
        \right), \quad
        S\left( \mbox{\scriptsize{$
\begin{array}{c|ccc}
		&&a+1&\\
	   \lam& a&& a\\
		&&a-1&
\end{array}  $}}
        \right)
\ee
                contain similar trigonometric factor
                besides
        $ S_{0}(\lam) $.

		Thus the antiferromagnetic spin
	$s$
		chain has very remarkable
		excitations. They are particles with
		rapidity and spin 1/2, but
                also kinks, relating the ``local vacua'',
		labeled by
	$a=0,\ldots, 2s$.
		Only transitions between the adjacent vacua
		are allowed in the
		physical Hilbert space. Some analogy with
		the Landau--Ginsburg
		picture in the topological field theory
		is evident, but not
		explored yet. On this intriguing note I
		finish the general
		discussion of the XXX${}_{s}$ model.

\section{XXX${}_{s}$ spin chain. Applications to the
		physical systems}

		The appearance of a parameter
	$s$
		in our disposal allows to use
		it to construct some models, going beyond
		the usual spin chains.
		Of particular interest is the limit of infinite
		spin
	$s\to\infty$,
		combined with the formal continuous limit
	$\Dl\to0$,
		which can be realized in many variants.
		I shall show,
		that such representative models of quantum
		field theory as
                Nonlinear Schroedinger equation (NLS) and $\Spher^2$
		nonlinear
	$\si$-model
		can be obtained from the XXX${}_{s}$ chain in
		this limit. The first model is nonrelativistic
		and of prime interest
                in the condensed matter physics, but the
		second one is
		interesting model of relativistic field theory.

		I begin with the NLS model, and employ the
		realization of spin
		variables via the complex canonical
                variables from section
        \ref{sone}:
\be
                S^+_n=\chi^*_n(2s-\chi^*_n\chi_n) \, ; \;\;
                S^-_n=\chi_n \, ; \;\;
                S^3_n=\chi^*_n\chi_n-s
\ee
		and suppose, that
	$s$
		is some complex number.

		The canonical commutation relations for
	$\chi^*_n$, $\chi_n$
		assume the usual form
\be
                [\chi_m^{},\chi^*_n]=\dl_{mn} \;\; .
\ee
		If
	$\Dl$
		is a lattice spacing,
	$\chi_n$ and $\chi^*_n$
		have order
	$\Dl^{1/2}$ in all expressions, where they enter
		in the normal order, i.e. all
	$\chi^*_n$
		to the left of all
        $\chi_m$
		with the same
	$n$.

		The invariant
	$C_{n,n+1}$ of two adjacent spins
	$S_{n}$ and
	$S_{n+1}$
		is given by
\ba
        2 C_{n,n+1}=2\sum_\al
		S_n^{\al} S_{n+1}^{\al}=
        2 S^3_{n}S^3_{n+1}+S^-_nS^+_{n+1}+S^+_nS^-_{n+1}=
\nonumber \\
	=2s^2-2s(\chi^*_n-\chi^*_{n+1})(\chi_n-\chi_{n+1})
			-(\chi^*_n-\chi^*_{n+1})^2
		\chi_{n+1}\chi_n.
\ea

		The second term in the RHS looks satisfactory,
		in the formal
		continuous limit it leads to the quadratic
		form of derivatives
		of field
	$\chi(x),\chi^*(x)$
\be
	\chi_n=\Dl^{1/2}\chi(x),\quad
        \chi^*_n=\Dl^{1/2}\chi^*(x),\quad x=n\Dl
\label{chin}
\ee
		due to the prescription
\be
        \chi_{n+1}=\Dl^{1/2}
                (\chi(x)+\Dl\chi'(x)+O(\Dl^2))  \quad .
\ee
		However the last term looks bad and
		does not lead to the
                desired expressions
	$(\chi^*(x))^2(\chi(x))^2$
		characteristic of the NLS model.
		The remedy is to use
		equivalent variables
        $\psi^{*}_n$, $\psi_n$
		on the lattice
\be
        \psi_n=(-1)^n\chi_n,\quad
                \psi^*_n = (-1)^n\chi^*_n
\ee
		and consider
        $\psi^*_n,\psi_n$ as producing the field
        $\psi(x),\psi^*(x)$ in the continuous limit as in
        (\ref{chin}).
		In the new variables we have
\be
        2C_{n,n+1}=2s^2-2s(\psi^*_n+\psi^*_{n+1})
                (\psi_n+\psi_{n+1}) +
                (\psi^*_n+\psi^*_{n+1})^{2}\psi_{n+1}\psi_n.
\ee
		Now the second term in the RHS is bad, but
		we add and substruct
        $4s(\psi^*_n\psi_n+\psi^*_{n+1}\psi_{n+1})$
		to change it into
\ba
        2 C_{n,n+1}= 2s^2-4s
                (\psi^*_n\psi_n+\psi^*_{n+1}\psi_{n+1})+
\nonumber  \\
        +2s(\psi^*_{n+1}-\psi^*_n)(\psi_{n+1}-\psi_n)+
                (\psi^{*}_n+\psi^*_{n+1})^{2}\psi_{n+1}\psi_n
                \quad ,
\ea
		which in the continuous limit will contain
		only good densities
        $n(x)=\psi^*\psi$, $h_0(x)=\psi'^*\psi'$ and
        $h_1(x)=(\psi^*)^2(\psi)^2$.
		We introduce the operator
	$J$ via
\ba
        && J(J+1)=2s(s+1)+2C_{n,n+1}=
\nonumber \\
          &&    = 4s^2+2s-8s\Dl n(x)+
                2s\Dl^3 h_0(x)+4\Dl^2 h_1(x)
\label{JJone}
\ea
		and consider the limit
        $s\to\infty$, $\Dl\to0$, $s\Dl=g$,
		where
	$g$
		is a new fixed parameter. We see, that the
		relation
        (\ref{JJone})
		allows to obtain the asymptotics of
	$J$ in this limit
\be
        J=2s+\frac as+\frac b{s^2}+\ldots \quad .
\ee
		Substituting this into the expression
\be
        \HAM_{n,n+1}(J)=2 \left. \frac{d}{dz}
                \ln \Gamma(z) \right| _{z=1+J}
\ee
                from section
        \ref{seight},
                and using Stirling formula
		for the asymptotics of
        $\Gamma(z)$
		and formal rule
\be
	\Dl\sum_n=\int dx
\ee
		we get for hamiltonian the expression
\be
        \HAM_{\hbox {\scriptsize lattice}}= \hbox{const} -\frac 1s N+
                \frac{g^2}{s^3}
                        \HAM^{\hbox {\scriptsize NLS}}+\ldots \quad ,
\ee
		where
\be
        N=\int \psi^*\psi dx
\ee
\be
        \HAM^{\hbox {\scriptsize NLS}}=\int
                \left[
                |\psi'(x)|^2+g(\psi^*)^2(\psi)^2
                \right]dx \;\; .
\ee
		This agrees nicely with the dispersion
		law for the energy of
		quasiparticles
\be
	h(\lam)=\frac s{\lam^2+s^2}=
                \frac 1s-\frac{\lam^2}{s^3}+\ldots \quad ,
\ee
                if we assume, that the momentum
	$\pp$
		of NLS particle is
		connected with the rapidity of XXX
		particle by a simple scale
\be
                \pp = \lam/g \;\;\;.
\ee

		The state
	$\Om$
		plays the role of no particle state of the NLS
                model. The bound states survive in the limit
	$s\to\infty$
		when
	$g<0$.
		For
	$g>0$
		the physical  hamiltonian is usually taken
		in the form
	$\HAM-\mu N$,
		where
	$\mu$
		is a chemical potential. Here the problem
		of ground state reappears and Dirac sea
		of particles, created by
        $\psi^*(x)$
		from
	$\Om$
		is to be used. The analogous problem
		for the original magnetic chain
		consists in the inclusion of
		magnetic field into hamiltonian
\be
        \HAM \to \HAM - \mu S^3  \;\; .
\ee
		This changes the picture of Dirac sea:
		only quasiparticles with
		rapidities, confined to some finite
		interval
        $-B \leqslant \lam \leqslant B$
		form the Dirac sea. The excitations are
		holes and quasiparticles with
        $|\lam|\geqslant B$.
		I shall not treat this case in any detail and
		finish the discussion of the NLS model.

		I turn to the second example --- the nonlinear
	$\si$-model with
		the field
	$n(x)$
		taking values in the 2-sphere
        $ \Spher^2$.

		The classical field
	$n(x)$
		can be described as a vector
	$n=(n_1,n_2,n_3)$
		subject to constraint
\be
		(n,n)=n^2_1+n^2_1+n^2_3=1
\label{constr}
\ee
		for all
	$x$.
		The canonical conjugate variable may
		be taken in
		the form
\be
        l=\frac 1\ga \, \partial_0 n\times n,\quad (l,n)=0.
\ee
		The canonical Poisson brackets
\ba
        \{l^\al(x),l^\bet(y)\}&=&\veps^{\al\bet\ga}
                l^{\/\ga}(x)\dl(x-y) \;\; ;
\label{lxly}
                \\
        \{l^\al(x),n^\bet(y)\}&=&\veps^{\al\bet\ga}
                n^{\ga} (x) \dl(x-y) \;\; ;
                \\
	\{n^\al(x), n^\bet(y)\}&=&0
\label{nxny}
\ea
		and hamiltonian
\be
        \HAM=\frac{\ga}{2}\int \left(
                l^2+\frac{n'^{2}}{\ga^2}\right)dx
\label{Hamint}
\ee
		follow from the lagrangian
\be
        {\cal L}= \frac1{2\ga} \,
                \partial_\mu n \, \partial_\mu n
\label{Lagr}
\ee
		and constraint
        (\ref{constr})
		in a usual way. Here
	$\ga$
		is a coupling constant, which is
		relevant only in quantum case.

		If we regularize the model going to
                the chain and introducing
		the variables
	$n_k, l_k$ as follows,
\be
        l_k=\Dl l(x),\quad n_k=n(x),\quad x=k\Dl \; ,
\label{deltalx}
\ee
		the brackets
        (\ref{lxly})--(\ref{nxny})
		will assume the ultralocal form
\ba
        \{l^\al_m,l^\bet_n\}&=
                &\veps^{\al\bet\ga}l^\ga_m\dl_{mn} \;\; ;
\label{lmln}  \\
        \{l^\al_m,n^\bet_n\}&=
                &\veps^{\al\bet\ga}n^\ga_m\dl_{mn} \;\; ;
\label{lmnn}  \\
        \{n^\al_m,n^\bet_n\}&=&0 \quad \quad .
\ea
		For each lattice site
	$m$
		the phase space is an orbit of the
		group
        $ \Ethree $
		of motions of
	$\RR^3$
                corresponding to the choice of Casimirs
\be
		n^2=1; \quad (n,l)=0 \;\; ,
\label{nlzero}
\ee
		which is also cotangent bundle of
        $\Spher^2$.
		Corresponding
		quantum Hilbert space can be realized as
        $L_2(\Spher^2)$.

		To make contact with the spin chains I
		mention, that this
                Hilbert space can be realized as an
		infinite spin limit of a
		Hilbert space of a pair of spin variables
		of spin
	$s$.
		Indeed,
                comparing the Klebsch--Gordan decomposition,
		already mentioned
		above,
\be
	D_s\otimes D_s=\sum^{2s}_{j=0} D_j
\ee
		and decomposition of
        $L_2(\Spher^2)$
\be
        L_2(\Spher^2)=\sum^\infty_{j=0} D_j   \;\; ,
\ee
		we see that
\be
        L_2(\Spher^2)=\lim_{s\to\infty} D_s\otimes D_s \;\;.
\ee
		Thus the
	$\si$-model variables
	$n_k,l_k$
		must be realized
		through the pair of spin variables
	$S_{2k-1}, S_{2k}$.
		The most naive way
\ba
        l_k & = & S_{2k-1}+S_{2k} \;\; ; \\
        n_k & = & \frac1{2s}(S_{2k}-S_{2k-1})
\ea
		or inversely
\ba
        S_{2k-1} & = & \frac12 l_k-sn_k \;\; ;
\label{Sodd}
                \\
        S_{2k} & = & \frac12 l_k+sn_k
\label{Seven}
\ea
		works. Indeed, from the spin commutation
		relations we reproduce
		the first two relations
        (\ref{lmln}), (\ref{lmnn})
		(in their quantum form) exactly and have
\be
        [n^\al_m , n^\bet_n]=
          \frac i{4s^2}\veps^{\al\bet\ga}l^\ga \dl_{m,n}
\ee
		with RHS vanishing for
	$s\to\infty$.
		Furthermore we have
\be
        (l_k,n_k)=0 \; ;\quad 4s^2 n^2_k+l^2_k=4s(s+1) \quad ,
\ee
		which reproduce
        (\ref{nlzero})
		in the limit
	$s\to\infty$.

                Now the idea is evident: we are
		to consider the XXX${}_{s}$ chain on
		the lattice of even length and take
                two adjacent points
	$(2k-1,2k)$
		as one lattice point for
	$\si$-model.
		Let us see,
		how the hamiltonian
        (\ref{Hamint})
		appears in the formal continuous limit.
		For that we are to
		estimate the invariants
	$C_{2k-1,2k}$ and $C_{2k,2k+1}$.
		We have
\be
	C_{2k-1,2k}=\frac14 l^2_k-s^2 n^2_k=
			\frac12 l^2_k-s(s+1)
\ee
		and
\be
	C_{2k,2k+1}=\frac14 l_kl_{k+1}+\frac s2
		(n_kl_{k+1}-n_{k+1}l_k) -
		s^2n_k n_{k+1}
\ee
		Now we estimate the invariant in
	$\Dl$
		expansion using the convention
        (\ref{deltalx})
		and
\be
	n_{k+1}=n(x)+\Dl n'(x)+\frac12 \Dl^2 n''(x)+\ldots
\ee
		to get
\ba
        2 C_{2k-1,2k}+2s(s+1) & = & \Dl^2
                l(x)^{2}     \quad ;
         \\
        2 C_{2k,2k+1}+2s(s+1) & = &
                        \Dl^2\left( l(x)^2-
                2s(n'(x),l(x)) - s^2(n''(x),n(x))
                \right) .
\nonumber
\ea
		Corresponding operators
        $J_{2k-1,2k}$ and $J_{2k,2k+1}$
		have order
	$\Dl^2$
		so that
\be
	\HAM_{2k-1,2k}+\HAM_{2k,2k+1}=
		2\psi(1+J_{2k-1,2k})+2\psi(1+J_{2k,2k+1)})
\ee
		can be easily calculated. Taking the sum over
	$k$
		with our usual understanding
\be
	\Dl\sum_k=\int dx
\ee
  and integrating by parts we get for
		the hamiltonian
	$\HAM^\si$
\be
        \HAM^\si=\frac1{2s\Dl\psi'(1)}
                \HAM^{\hbox {\scriptsize XXX}}
                =\frac1{s} \int \left(\left(l-\frac12
                sn'\right)^2+\frac{s^2}4(n')^2\right)dx  \; ,
\ee
		which turns into
        (\ref{Hamint})
                if we comment, that the map
\be
                l  \to  l+\al n' \; , \quad
                n  \to  n
\label{theshift}
\ee
		for any
	$\al$
		is a canonical transformator for the brackets
        (\ref{lxly})--(\ref{nxny}).
		The coupling constant
	$\ga$
		is connected with spin $s$ as follows
\be
		\ga=\frac 2s \;\;.
\ee
		The shift
        (\ref{theshift})
		is produced by a topological term,
		added to the lagrangian
        (\ref{Lagr}),
\be
        {\cal L}_\vartheta=\frac\vartheta{8\pi}(\partial_\mu n
		\times\partial_\nu
                n,n)\veps_{\mu\nu} \; .
\ee
		Indeed, such addition changes only
		the definition of the
		canonical momenta
\be
        l \to l+\frac\vartheta{4\pi} n'
\ee
		and thus we can interpret the model
		obtained from XXX chain as
		nonlinear $\si$-model with
        $\vartheta$-term with
\be
                \vartheta=2\pi s \;\; .
\ee
		The
        $\vartheta$-term with
        $\vartheta=2\pi n$
		is trivial; in other words
        $\vartheta$ is defined mod $2\pi$.
                Thus we see an important difference
                of integer and half--integer spin
        $s$
		used in our construction.
		The topological term is present
                only if spin is half--integer.
                This phenomenon and its consequences
                are discussed in
                detail by Professor I.~Affleck
		in his Les--Houches lectures of 1993 and
		I can only refer you to the
		corresponding proceedings.

                Unfortunately the described connection
                was not yet realized in any
		real computation for the nonlinear
	$\si$-model.
		In particular we
		expect, that the magnetic chain in the
		relevant thermodynamic
		limit must have an excitation of spin 1
		with relativistic
		dispersion law
\ba
                \pp(\lam) & = & c\, \sinh\lam \quad ; \\
                \ep(\lam) & = & c\, \cosh\lam
\ea
		with parameter
		$c$ being exponentially small for
	$s\to\infty$
\be
        c=e^{-s/2} \;\; .
\ee
		This will lead to the realizations of
		dimensional
                transmutation program giving mass
	$m$
		via the lattice spacing $\Dl$
		and the coupling constant
	$\ga=2/s$
\be
        m=\frac1\Dl e^{-1/\ga}   \quad .
\ee
		Apparently to achieve this we are to
		understand which portion of
		the infinite tensor product of the
		spin chain is compatible with
		the formal continuous limit we just described.

		One comment could be relevant to this
		program. It is natural to
		combine the adjacent Lax operator
		into product
        $L_{2k,a}(\lam)
                L_{2k-1,a}(\lam)$
		and perform the change of variables
        (\ref{Sodd}), (\ref{Seven})
		there. It turns out, that this new Lax
		operator has a new
		local vacuum in
        $L_2(\Spher^2)$
		and the corresponding BAE
		look like
\be
	\left(\frac{\lam_k-is}{\lam_k+is}\,
		\frac{\lam_k+i(s+1)}
		{\lam_k-i(s+1)}
	\right)^{N/2}=
		\prod_{j\neq k}
		\frac{\lam_k-\lam_j-i}{\lam_k-\lam_j+i} \; .
\label{BAEcorr}
\ee
                The alternating value of spin necessary
		to maintain our conventions
        (\ref{deltalx})
		is  manifest here. However the proper
		choice of solutions to
                these equations is not done yet.
		I leave it as a challenge and
		stop here the discussion of nonlinear
	$\si$-model.

		I also finish considerations of
		XXX chains (with one revisit
                in section
        \ref{stwelve})
                and turn to the XXZ chains.

\section{ XXZ model}
\label{sten}

                As was told above, this model is a deformation of XXX
		model with one new
                parameter. We shall denote this parameter by
	$q$
		or
	$\ga$ with $q=e^{i\ga}$.
                The deformation uses the
	$q$-analogues
		of usual number
\begin{equation}
        [x]_q=\frac{q^x-q^{-x}}{q-q^{-1}}=
           \frac{\sin\ga x}{\sin\ga}=
              \prod_{n=-\infty}^{\infty}
                \left( \frac{x+n\pi\ga^{-1}}
                        {1+n\pi\ga^{-1}} \right) \;\; ,
\label{iksq}
\end{equation}
                which effectively change the complex plane to a strip
		via the
                multiplicative averaging in
                (\ref{iksq}).
                We begin with the construction of the Lax operator of
		XXZ model using the
                matrix analogue of this averaging. For classical
                spin variables the formal averaging
\begin{equation}
                L^{\hbox {\scriptsize XXZ}}_{n,a}(\lam)=
                        \prod^\infty_{k=-\infty}
                L_{n,a}^{\hbox {\scriptsize XXX}}(\lam+ik\pi\ga^{-1})
\end{equation}
                can be evaluated to lead to the expression
\begin{equation}
                L^{\hbox {\scriptsize XXZ}}_{n,a}(\lam)=\frac1{\sin\ga}
                \left(\begin{array}{cc} \sinh\ga(\lam+i\tilde S^3_n)&
                \sin\ga\tilde S_n^-\\
		\sin\ga\tilde S^+_n
                &\sinh\ga(\lam-i \tilde S^3_n)\end{array}\right) \quad,
\end{equation}
                where
	$\tilde S^3_n$, $\tilde S^{\pm}_n$
		are some function of the original spin
                variables
	$S^3_n$,
	$S^{\pm}_n$
                of XXX model. Taking this as heuristic
                consideration we shall look for the Lax operator
        $L_{n,a}(\lam)$
		in the form
\begin{equation}
                L_{n,a}(x)=\left(
		\begin{array}{cc}
                        x q^{S^3_n}- x^{-1} q^{-S^3_n} &
                        (q-q^{-1})S^-_n\\
                        (q-q^{-1})S^+_n &
                        xq^{-S_{n}^{3}}-x^{-1}q^{S_{n}^{3}}
		\end{array}
                \right) \quad ,
\label{Laxchi}
\end{equation}
                using multiplicative spectral parameter
\begin{equation}
                x = q^{-i\lam}
\end{equation}
                and quantum operator
	$q^{S^3_n}$,
	$S^\pm_n$.
		We shall check, that
        $L_{n,a}(x)$
		satisfy the FCR
\begin{equation}
        R_{a_1,a_2}\left(x/y\right)
           L_{n,a_1}(x)L_{n,a_2}(y)=
              L_{n,a_2}(y)L_{n,a_1}(x)
                R_{a_1,a_2}\left(x/y\right) \quad ,
\label{FCRZ}
\end{equation}
                where
        $ R $
		is a
	$q$-deformed
                analogue of the
        $ R $-matrix
                from section
        \ref{sfour}
                (see (\ref{Rfour}))

\begin{equation}
                R=\left(\begin{array}{cccc}
                        a&&&\\
                        &b&c&\\
                        &c&b&\\
                         &&&a
		\end{array}\right)
\end{equation}
                with
\begin{equation}
        a=qx-q^{-1}x^{-1}; \ \ \ b=x-x^{-1}; \ \ \ c=q-q^{-1}.
\end{equation}
                To check FCR
                (\ref{FCRZ})
                it is convenient to twist it a little, introducing
\begin{eqnarray}
        \tilde L_{n,a}(x) & = & Q(x)L_{n,a}(x)Q^{-1}(x) \;\; , \\
        \tilde R_{a_1,a_2}(x) & = & Q(x)\otimes Q(y) \
                 R_{a_1,a_2}(x/y) \ Q^{-1}(x)\otimes Q^{-1}(y) \; ,
\end{eqnarray}
                where
	$Q(x)$
		is a matrix in the auxiliary space
\begin{equation}
                Q(x)=\left(
		\begin{array}{cc}
			x^{1/2} 	&0\\
			0& 		x^{-1/2}
		\end{array}
		\right).
\end{equation}
                It is clear, that FCR is true for original
	$L_{n,a}(x)$
		if it holds for
        $\tilde L_{n,a}(x)$.

                Now observe, that
	$\tilde L_{n,a}(x)$
		and
        $\tilde R_{a_1,a_2}(x)$
		have a simple
	$x$-dependence
\begin{equation}
                L=xL_+-x^{-1}L_- \; , \quad
                R=xR_+-x^{-1}R_- \;\; ,
\label{Ltilda}
\end{equation}
                where we dropped indeces
        $n$, $a_1$, $a_2$, $ \tilde {} $,
		and matrices
        $L_+$, $L_-$, $R_+$, $R_-$
		are given by
\begin{equation}
        L_+=\left(
		\begin{array}{cc}
                        q^{S^3}&        (q-q^{-1})S^-\\
                        0&              q^{-S^3}
		\end{array}\right);\quad
                L_-=\left(
		\begin{array}{cc}
                        q^{-S^3}       &        0\\
                        -(q-q^{-1})S^+ &     q^{S^3}
		\end{array}
                \right)  \quad ;
\label{LpLm}
\end{equation}
\be
                R_+ = \left(
		\begin{array}{cccc}
                            q &         &                 &\\
                              &   1     &      q-q^{-1}   &\\
                              &         &      1          &\\
                              &         &                 &q
		\end{array}
                \right)         \quad ;
\ee
\be
                R_- = \left(
		\begin{array}{cccc}
                        q^{-1} &                &                 &\\
                               &  1             &      0          &\\
                               &  -(q-q^{-1})   &      1          &\\
                               &                &                 & q^{-1}
                \end{array}\right)   \quad .
\ee
                It is clear, that
\begin{equation}
                R_-=\PP R^{-1}_+\PP
\end{equation}
                and
\begin{equation}
                R_+-R_-=\left(q-q^{-1}\right)\PP \;\; ,
\label{RpRm}
\end{equation}
                where
        $ \PP $
		is a permutation matrix.

                Separation of spectral parameters in FCR shows,
                that it is true, if the
                following 7 relations hold:
\begin{equation}
                R L^1_\pm L^2_\pm=L^2_\pm L^1_\pm R \;\; ,
\label{RLLpm}
\end{equation}
                where
        $R$
		can be
        $R_+$
		and
        $R_-$
                and labels 1 and 2 substitute
	$a_1$
                and
	$a_2$;
		furthermore
\begin{eqnarray}
        R_+L^1_+L^2_- & = & L^2_-L^1_+R_+ \;\; ;
\label{RLLp}
                \\
        R_-L^1_-L^2_+ & = & L^2_+L^1_-R_-
\label{RLLm}
\end{eqnarray}
                and
\begin{equation}
                R_+L^1_-L^2_+-R_-L^1_+L^2_-=
                        L^2_+L^1_-R_+-L^2_-L^1_+R_-.
\label{RLLmin}
\end{equation}
                Only three of them are independent and we
                take as such two of the relation
                (\ref{RLLpm})
                for
	$L_+$
		and
	$L_-$
		separately and relation
                (\ref{RLLp}).
                The two other relations in
                (\ref{RLLpm})
                and relation
                (\ref{RLLm})
                are easily reduced to the chosen ones if one applies
		the permutation taking into account, that
\begin{equation}
                L^1 = \PP L^2 \PP \;\; .
\end{equation}
                Finally relation
                (\ref{RLLmin})
                is checked if one uses the property
                (\ref{RpRm})
                to substitute unwanted
        $R_+$
		by
        $R_-$
		and vice versa.

                Now it is easy to check, that the basic
                relations, say
\begin{eqnarray}
                R_+L^1_+L^2_+ & = & L^2_+L^1_+R_+ \;\; ;
\label{basicone}
        \\
                R_+L^1_-L^2_- & = & L^2_-L^1_-R_+ \;\; ; \\
                R_+L^1_+L^2_- & = & L^2_-L^1_+R_+
\label{basictoo}
\end{eqnarray}
                are satisfied, if
        $q^{S^{3}}$,
        $S^{+}$
		and
        $S^{-}$
		satisfy the commutation
                relations of the
	$q$-deformed
        $ \sltwo$
                from section
        \ref{stwo}.
                This
                finishes the proof of the FCR
                (\ref{FCRZ}).
                It is worth to mention, that the XXX${}_{s}$
                Lax operator
                (\ref{Laxs})
                is obtained from
                (\ref{Laxchi})
                in the limit
	$q\to1$,
		so FCR is proved also for it.

                Another comment is, that we can consider the relations
                (\ref{basicone})--(\ref{basictoo})
                together with the structure
                (\ref{LpLm})
                as alternative definition of the
	$q$-deformed
        $\slqtwo$
                algebra, which will prove to be
                convenient in what follows.

                If we renormalize
        $R_\pm$
		by
\begin{equation}
                R_+\to q^{-1/2}R_+,\quad
                R_-\to q^{-1/2}R_-
\end{equation}
                we see that they take form
                (\ref{LpLm})
                where
        $q^{S^3}$
		and
        $S^{\pm}$
		realize the 2-dimensional representation
\begin{equation}
                q^{S^3}=\left(
		\begin{array}{cc}
			q^{1/2} 	& 0\\
			0		& q^{-1/2}
		\end{array}
		\right), \ \ \
                S^+=\left(
		\begin{array}{cc}
			0	&q^{1/2}\\
			0	&0
		\end{array}
		\right), \ \ \
                S^-=\left(
		\begin{array}{cc}
			0		&0\\
			q^{-1/2}	&0
		\end{array}
                \right),
\end{equation}
                deforming the Pauli matrices.

                The
        $R$-matrix
                and Lax operators above can be
                considered as representation
                of the  universal
	$\RMA$-matrix,
                which in this case corresponds to the
	$q$-deformed
                affine algebra
        $\slaff$.

                In some sense this algebra (see a lot about it in
		Professor Miwa lectures) is
                more natural object than Yangian, which is a special
		contraction of it
                when
	$q\to1$
		(or
	$\ga$
		tends to 0) in conjunction with a renormalization
                of the grading parameter
        $\lam$.

                The
	$q$-deformed
        $\sltwo$
		has finite dimensional
                representations, analogous
                to those in the nondeformed case;
		their dimension is
	$2s+1$,
		where spin
        $ s $
                is integer or half--integer. Besides it has other
                interesting representations, called cyclic and already
		briefly mentioned in
                section
        \ref{stwo};
                these representations have no limit, when
	$q\to1$.

                The deformed representations in $\CC^{2s+1}$ are of
		highest weight type, so
                that there is a state
	$\om$
		such, that
\be
                S^+\om = 0 \; , \quad
                q^{S^3}\om = q^s\om \;\; .
\ee
                This allows to repeat all constructions of Algebraic
		Bethe Anzatz and the BAE assume the form
\begin{equation}
                \frac   {\sinh(\lam_k+is\ga)}
                        {\sinh(\lam_k-is\ga)}
		=
		\prod_{j\ne k}
                \frac   {\sinh(\lam_k-\lam_j+i\ga)}
                        {\sinh(\lam_k-\lam_j-i\ga)}   \quad ,
\label{BAEZ}
\end{equation}
                where we returned to the additive spectral parameter
        $\lam$
		to make
                (\ref{BAEZ})
                look rather similar to BAE
                (\ref{BAEs}).

                The investigation of these equations is to large extent
		similar to what was
                done before in XXX case. One encounters
                complexes of roots,
		one can estimate
                the number of Bethe vectors etc. A new feature
                is a role, played by the
                arithmetic nature of
	$\ga$.
		For instance if
	$\ga/\pi$
		is rational then
	$q$
                is a root of unity and all specifics of
        $\slqtwo$
                for such a case is to be
                taken into account. We have no time to discuss
                it in any detail.

                I turn now to the problem of constructing
                the local hamiltonian. For that
                we need to control better the invariant
                of the pair of spins.

                It is convenient to use the matrix
\begin{equation}
                L=L_+L_-^{-1}
\end{equation}
                to describe the deformed spins. The matrix
	$L^{-1}_-$
		is given by
\begin{equation}
                L^{-1}_-=\left(
		\begin{array}{cc}
                        q^{S^{3}}                 &0\\
                        (q-q^{-1})S^{+}       &q^{-S^{3}}
		\end{array}
		\right).
\end{equation}
                The contractions
	$q\to1$
		is especially transparent here with expansion
\begin{equation}
                L=I+2i\ga\left(
		\begin{array}{cc}
                        S^3     &S^-\\
                        S^+     &-S^3
		\end{array}
		\right)
						+\ldots
\end{equation}
                for
	$\ga\to0$,
		where
        $S^3$,
        $S^{\pm}$
		are nondeformed spin variables.

                The entries of matrix
	$L$
                satisfy the relations which can be cast in the
                matrix form
\begin{equation}
                L^1R^{-1}_-L^2R_-=R^{-1}_+L^2R_+L^1.
\end{equation}
                There exists the generalized trace (so called
	$q$-trace)
\begin{equation}
                \tr_qA=\tr(DA)
\end{equation}
                with
\begin{equation}
                D=\left(
                \begin{array}{lc}
			q^{-1}	&0\\
			0	&q
		\end{array}
                \right)  \;\; ,
\end{equation}
                such that for any matrix
	$A$
		we have
\begin{equation}
                \tr^2_qR^{-1}A^2R=\tr_q \Id^1 A \;\; ,
\end{equation}
                where we calculate
	$\tr_q$
		only over the second factor in the tensor
                product of two auxiliary spaces.

                From these properties we see immediately that
\begin{equation}
                [\tr_qL, \ L]=0  \;\; ,
\end{equation}
                i.e.,
	$\tr_qL$
		plays the role of Casimir and we shall denote it as
	$C$.

                The explicit calculation gives
\begin{equation}
                C=      \left(q+q^{-1}\right)
                        \left(q^{2S^3}+q^{-2S^3}\right)
			+
                        \left(q-q^{-1}\right)^2
                        \left(S^+S^-+S^-S^+\right) \;\; ,
\end{equation}
                or
\begin{equation}
                C=q^{2S^3+1}+q^{-2S^3-1}+
                2\left(q-q^{-1}\right)^2 S^-S^+ \;\; ,
\end{equation}
                so that in the irreducible representations of spin
        $s$
		the eigenvalues of
	$C$
		are given by
        $q^{2s+1}+q^{-2s-1}$.
		This prompts the introduction of the
                operator
	$J$
		such that
\begin{equation}
                C=q^{2J+1}+q^{-2J-1},
\label{CasimJ}
\end{equation}
                which is an analogue of
	$J$
		from the usual
        $\sltwo$.

                Armed with this knowledge we can attack
                the problem of local hamiltonian in
                a way analogous to that of section
        \ref{seight}.

                We shall take two spin operators and
                describe them by the corresponding
	$L$-operators,
		which we shall denote by
	$L_\pm$
		and
	$M_\pm$.
		Let
	$L(x)$
		and
	$M(y)$
		be the corresponding Lax operator, taken in the form
                (\ref{Ltilda}).
                The FCR we shall use to find the representation for the
        $ R^{s,s}(x) $
		Lax operator
        $ R(x) $
		looks as follows
\begin{equation}
                L\left(1/x\right)M\left(1/y\right)
                        R\left(x/y\right)
		=
                R\left( x/y\right)M\left(1/y\right)
                        L\left(1/x\right).
\end{equation}
                Putting
        $R\left(x\right)$ into the form
\begin{equation}
                R(x)=\PP r(x)
\end{equation}
                we have instead
\begin{equation}
                M\left(1/x\right)L\left(1/y\right)
                        r\left(x/y\right)=
                r\left(x/y\right)M\left(1/y\right)
                        L\left(1/x\right).
\label{MLr}
\end{equation}
                Altogether
                (\ref{MLr})
                comprises four matrix equations. Taking
	$r$
		to be a function of invariant of spin
	$S$
		and
	$T$
                we trivially satisfy two of them.
                One of remaining looks
                as follows
\begin{equation}
        \left(x^{-1} M_+L_-+xM_-L_+\right)r(x)
		=
        r(x)\left(xM_+L_-+ x^{-1} M_-L_+\right).
\end{equation}
                We shall take lower left element of this equation
\begin{equation}
        \left(x^{-1} q^{-T^3}S^++xq^{S^3}T^+\right)r(x)
		=
        r(x)\left(xq^{-T^3}S^++ x^{-1} q^{S^3}T^+\right)
\end{equation}
                and consider it in the subspace
\begin{equation}
                q^{T^3}S^++T^+q^{-S^3}=0 \;\; ,
\end{equation}
                where the invariant takes the form
        (\ref{CasimJ}) with
\begin{equation}
                J=T^3+S^3 \;\; .
\end{equation}

                We shall look for
	$r(x)$
		as a function of
	$q^{2J+1}$.
		Our equation looks as follows
\begin{equation}
        \left(x^{-1} q^{-J}-xq^J\right)q^{S^3}S^+r(x, q^{2J+1})
		=
        r(x,q^{2J+1})\left(xq^{-J}-x^{-1} q^J\right) q^{S^3}S^+.
\label{relat}
\end{equation}
                Using the commutation relation
\begin{equation}
                S^+q^{2J+1}=q^{-2}q^{2J+1}S^+
\end{equation}
                we transform the equation
        (\ref{relat})
                into the functional equation
\begin{equation}
                \frac	{r(qw)}
			{r(q^{-1}w)}
		=
		\frac	{1-x^2w}
                        {x^2-w}   \quad ,
\label{funleq}
\end{equation}
                where
\begin{equation}
                w=q^{2J+1}.
\end{equation}
                We shall not discuss this equation here;
                suffice to say, that it is a
	$q$-deformation
		of equation
                (\ref{functeq})
                and its solution is given by some
	$q$-deformation
		of
        $\Ga$-function
		(known
                also as
	$q$-exponent
                and exp of quantum dilogarithm). This finishes the
                general discussion of the XXZ model.

\section{Inhomogeneous chains and discrete time shift}
\label{seleven}

                Here I shall present some development of
                the general scheme of Algebraic
                Bethe Anzatz which will allow
                to include more dynamical models under its
                spell. Until now two main
                observables --- momentum and energy were treated
                differently. We used discretized space,
                but continuous time; thus we
                introduced a finite space shift
	$U$
		and infinitesimal generator of
                time shift
	$H$.
                To make our consideration more
                manifest invariant it is
                natural to discretize also time.
                Exactly this will be done here. We used
                lattice as a way of regularization and
                always had in mind corresponding
                continuous limit. But some people take
                discrete space-time more seriously
                as an inevitable consequence of gravity.
                I shall not open this discussion
                here sticking to the formal aspects only.

                The shift operator
	$U$
		was obtained above as a trace of the monodromy at
                some distinguished value of spectral parameter.
                We need now two such
                distinguished values. The way to achieve
                this goal is to consider a
                specially inhomogenous chain with the spectral
                parameter taking alternating
                values
        $\lam\pm\om$
		for some fixed
	$\om$.
		This simple idea indeed works as
                we shall see momentarily.

\begin{figure}
\begin{picture}(350,120)(0,0)
  \multiput(80,60)(40,0){6}{\line(-1,-1){35}}
  \multiput(80,60)(40,0){6}{\line(1,1){45}}
  \multiput(80,60)(40,0){6}{\line(-1,1){45}}
  \multiput(80,60)(40,0){6}{\line(1,-1){35}}


 \put(330,60){\vector(-1,0){15}}
 \put(330,60){\vector(1,0){15}}
 \put(330,60){\vector(0,1){15}}
 \put(330,60){\vector(0,-1){15}}
 \put(346,56){$E$}
 \put(304,56){$W$}
 \put(326,78){$N$}
 \put(326,35){$S$}

 \multiput(80,59)(0,2){2}{\line(-1,-1){20}}
 \multiput(80,59)(0,2){2}{\line(1,-1){20}}

 \multiput(120,59)(0,2){2}{\line(-1,-1){20}}
 \multiput(120,59)(0,2){2}{\line(1,-1){20}}

 \multiput(160,59)(0,2){2}{\line(-1,-1){20}}
 \multiput(160,59)(0,2){2}{\line(1,-1){20}}

 \multiput(200,59)(0,2){2}{\line(-1,-1){20}}
 \multiput(200,59)(0,2){2}{\line(1,-1){20}}

 \multiput(240,59)(0,2){2}{\line(-1,-1){20}}
 \multiput(240,59)(0,2){2}{\line(1,-1){20}}

 \multiput(280,59)(0,2){2}{\line(-1,-1){20}}
 \multiput(280,59)(0,2){2}{\line(1,-1){20}}

\end{picture}
\caption{ Discrete space time and initial saw }
\label{insaw}
\end{figure}
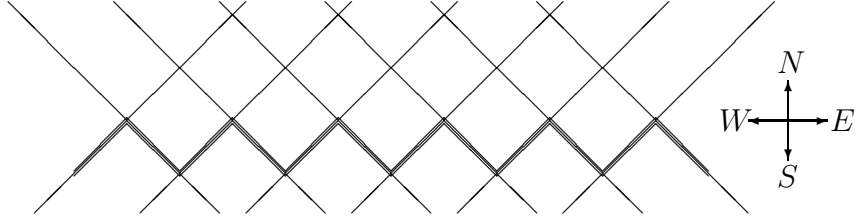

                Let
        $L_{n,f}(\lam)$
                be a fundamental Lax operator with the same
                quantum space
	$h_n$
		and auxiliary space
	$V_f$.
		The monodromy of the inhomogeneous
                chain is given by
\begin{equation}
                T_f(\lam,\om)=L_{2N,f}(\lam+\om)L_{2N-1,f}(\lam-\om)
                \ldots L_{2,f}(\lam+\om)
                L_{1,f}(\lam-\om).
\end{equation}
                We want to argue, that
\begin{equation}
                U_+ = \tr_{f}T_f(\om,\om),\quad
                        U_- = \tr_{f}T_f(-\om,\om)
\end{equation}
                play the role of shifts in the characteristic
                (light-like) directions in the
                discrete space-time. This space-time is natural to draw
                as shown in Figure
        \ref{insaw},
                where the
                monodromy is taken as a product along the initial saw,
		depicted
                by a fat line.

                The space runs from east to west and
                time from south to north. The
        $L_{2n,f}(\lam+\om)$
		is transport along
	$NW$
                direction and
        $L_{2n+1,f}(\lam-\om)$
                is the same in
	$SW$
		direction.

                To see the shift properties of operators
	$U_\pm$
		we consider even more
                inhomogeneous monodromy inserting into
                the string of Lax operators
        $L_{n,f}(\lam \pm \omega)$
                the Lax operator
        $L^{-1}_{f,a}(\mu-\lam)$,
		treating the space
	$a$
		as quantum
                and
	$f$
		as auxiliary. This admixture is put between
        $L_{2n,f}(\lam+\om)$
		and
        $L_{2n-1,f}(\lam-\om)$.
		The new monodromy looks as follows
\begin{eqnarray}
    &&    T_{f}(\lam,\om | a,n,\mu)=
        L_{2N,f}(\lam+\om)L_{2N-1,f}(\lam-\om)
                        \ldots
        \\
       &&     \ldots L_{2n,f}(\lam+\om)L^{-1}_{f,a}(\mu-\lam)
                L_{2n-1,f}(\lam-\omega)\ldots
                L_{2,f}(\lam+\om)L_{1,f}(\lam-\om).
\nonumber
\end{eqnarray}
                Due to the similar FCR for all entries
                here this monodromy also satisfies
                FCR and its trace
\begin{equation}
                F(\lam,\om | a,n,\mu)=\tr_f T_f(\lam,\om| a,n,\mu)
\end{equation}
                gives a commuting family as function of
        $\lam$.

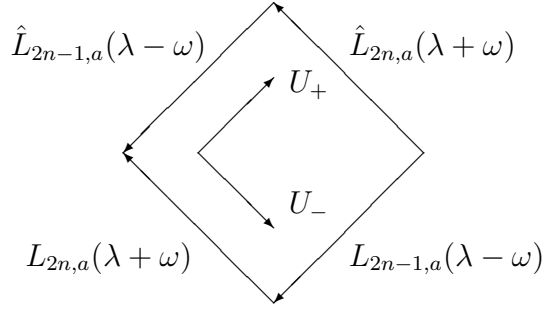
\begin{figure}
\setlength{\unitlength}{1mm}
\begin{center}
\begin{picture}(100,40)

\put (50,0){\vector(-1,1){20}}
\put (50,40){\vector(-1,-1){20}}
\put (70,20){\vector(-1,1){20}}
\put (70,20){\vector(-1,-1){20}}
\put (40,20){\vector(1,1){10}}
\put (40,20){\vector(1,-1){10}}
\put (52,28){$U_{+}$}
\put (52,12){$U_{-}$}
\put (17,5){$L_{2n,a}(\lambda+\omega)$}
\put (15,33){$\hat{L}_{2n-1,a}(\lambda-\omega)$}
\put (60,33){$\hat{L}_{2n,a}(\lambda+\omega)$}
\put (60,5){$L_{2n-1,a}(\lambda-\omega)$}
\end{picture}
\end{center}
\caption{ Zero curvature equation }
\label{zerocurv}
\end{figure}

                Now let us look at
        $F(\lam,\om| a,n, \mu)$
		at distinguished points
        $\lam=\pm\om$.
		We use the normalization
\begin{equation}
                L_{n,f}(0) = \PP_{n,f} \quad ,
\end{equation}
                where
        $\PP$
		is a permutation of
        $\hh_n$
		and
        $\VV_f$,
		which coincide as
                spaces. We have
\ba
      &&  F(\om,\om| a,n,\lam)=
                 \tr_f\left(L_{2N,f}(2\om)P_{2N-1,f}\ldots
        \right.
\nonumber \\
         &&  \left.
             \ldots L_{2n,f}(2\om)
                L^{-1}_{f,a}(\lam-\om)P_{2n-1,f}\ldots
                        L_{2,f}(2\om)P_{1,f}
		\right).
\ea
                Note, that we changed parameter
	$\mu$
		to
        $\lam$.
		Due to relation
\begin{equation}
        L^{-1}_{f,a}(\lam-\om)\PP_{2n-1,f}=
                \PP_{2n-1,f}L^{-1}_{2n-1,a}(\lam-\om)
\label{LPPL}
\end{equation}
                we can bring
        $L^{-1}_{f,a}(\lam-\om)$
		to extreme right to get
\begin{equation}
        F\left(\om,\om| a,n,\lam\right)=U_+L^{-1}_{2n-1,a}(\lam-\om).
\label{FUL}
\end{equation}
                Analogously
\begin{equation}
        F\left(-\om,\om| a,n,\lam\right)=L^{-1}_{2n,a}(\lam+\om)U_-.
\label{FLU}
\end{equation}
                The commutativity of
	$F$
		in the LHS of
                (\ref{FUL})
                and
                (\ref{FLU})
                leads to the equation
\be
        U_+ L^{-1}_{2n-1,a}(\lam-\om)
                L^{-1}_{2n,a}(\lam+\om)U_-=
        L^{-1}_{2n,a}(\lam+\om)U_-
                U_+L^{-1}_{2n-1,a}(\lam - \om) \; ,
\ee
                which we can rewrite due to commutativity of
	$U_+$
		and
	$U_-$
		in the form
\begin{equation}
                L_{2n,a}(\lam+\om)L_{2n-1,a}(\lam-\om)=
                     U_-L_{2n-1,a}(\lam-\om)
			U^{-1}_- \ U^{-1}_+
                L_{2n,a}(\lam+\om)U_+.
\end{equation}
                This equation has natural interpretation
                as a zero curvature condition for
                the transport around the elementary plaquette
                in our space-time. On Figure
        \ref{zerocurv}
                we use notations
\ba
        \widehat L_{2n-1,f}(\lam-\om) & = &
                U_-L_{2n-1,f}(\lam-\om)U^{-1}_- \quad ; \\
        \widehat L_{2n,f}  (\lam+\om) & = &
                U^{-1}_+ L_{2n,f}(\lam+\om)U_+ \; .
\ea
                Thus we see, that
	$U_+$
                is a shift in
        $NE$
                direction and
	$U^{-1}_-$
		is shift in
	$NW$
		direction. In terms of shifts in
	$N$ (time)
		and
	$W$ (space)
		directions
	$e^{-iH}$
		and
	$e^{-iP}$ we have
\begin{equation}
        U_+=e^{-i(H-P)/2}\; , \quad U_-=e^{i(H+P)/2} \; .
\label{UpUm}
\end{equation}
                This is our main assertion.
                Now I shall describe a more explicit
                expressions for
	$U_\pm$.
		Using our usual Ansatz
\begin{equation}
                L_{n,f}(\lam)=\PP_{n,f}l_{n,f}(\lam)
\end{equation}
                we get
\begin{equation}
        U_+=\tr_f\left(\PP_{2N,f}l_{2N,f}(2\om)\PP_{2N-1,f}\ldots
              \PP_{2,f}l_{2,f}(2\om)\PP_{1,f}\right)
\end{equation}
                and bringing all
	$l_{n,f}(2\om)$
		to right we get
\begin{equation}
                U_+=V\prod_n l_{2n,2n-1}(2\om)  \;\; ,
\end{equation}
                where
        $V^{-1}$
		is a shift
	$n\to n+1$.

                Analogously we have
\begin{equation}
                U_-=\prod_n l_{2n,2n-1}(-2\om)V \;\; ,
\end{equation}
\begin{equation}
                l_{2n,2n-1}(-\lam)=l^{-1}_{2n,2n-1}(\lam) \quad .
\end{equation}
                From identification (\ref{UpUm}) we get
\be
                e^{iP}=U_+U_-=V^2    \quad ;
\ee
\ba
          e^{-iH}&=&U_+U^{-1}_-=V\prod l_{2n,2n-1}(2\om)V^{-1}
                \prod l_{2n,2n-1}(2\om)=
\nonumber \\
      & = & \prod l_{2n+1,2n}(2\om)\prod l_{2n,2n-1}(2\om) \;\; .
\ea
                We see, that the physical space shift is shift
	$n\to n+2$;
                in other words two
                lattice sites of our space lattice
                constitute one physical site. We already
                have seen such trick in the discussion of nonlinear
	$\si$-model.

                The derivation of the BAE for the inhomogeneous
                chain does not differ from
                that given above. The equations look like
\begin{equation}
                \left(
                        \frac   {\al(\lam_j+\om)\al(\lam_j-\om)}
                                {\dl(\lam_j+\om)\dl(\lam_j-\om)}
		\right)^N
                =
                \prod_{k\ne j}S(\lam_j-\lam_k) \;\; ,
\label{BAEch}
\end{equation}
                where
        $\al(\lam)$
		and
        $\dl(\lam)$
		are local eigenvalues and factor
        $S(\lam)$
                in the RHS is a quasiparticle phase factor.
                From this we
                read the quasiparticle momentum and energy:
\begin{equation}
                e^{ip}=\frac{\al(\lam+\om)\al(\lam-\om)}
                        {\dl(\lam+\om)\dl(\lam-\om)};\quad
                e^{ih}=\frac{\dl(\lam+\om)\al(\lam-\om)}
                        {\al(\lam+\om)\dl(\lam-\om)} \quad .
\end{equation}
                These expressions nicely turn into our
                previous formulas in the limit
	$\om\to0$.
                This finishes our description of a general
                scheme and I turn to
                examples.

\section { Examples of dynamical models
                in discrete space-time }
\label{stwelve}

                I shall present two examples for the
                general scheme of the previous
                section. The first is associated with
        XXX${}_s$
		spin chain. The BAE
\begin{equation}
                \left(
                        \frac   {(\lam_j+\om+is)(\lam_j-\om+is)}
                                {(\lam_j+\om-is)(\lam_j-\om-is)}
                \right)^N
		=
                \prod^l_{k\ne j}\frac   {\lam_j-\lam_k+i}
                                        {\lam_j-\lam_k-i}
\end{equation}
                can be investigated as above. The spectrum
                around the antiferromagnetic
                state, already described in section
        \ref{seight},
                leads to the dispersion law for
                the excitations
\begin{eqnarray}
        \pp(\lam) & = & \arctg \sinh \pi(\lam +\omega) +
                \arctg \sinh \pi(\lam - \omega) \; ; \\
        \ep(\lam) & = & \frac{1}{\cosh \pi(\lam +\omega)} +
                        \frac{1}{\cosh \pi(\lam -\omega)} \; .
\end{eqnarray}
                In the limit
	$\om\to\infty$
		we get the relativistic one particle spectrum
\begin{equation}
                \pp(\lam)=m \, \sinh \pi\lam \; , \quad
                \ep(\lam)=m \, \cosh \pi\lam \; ,
\end{equation}
                where
	$ m$
		is obtained via the ``dimensional transmutation''
\begin{equation}
                m=\frac1\Dl e^{-\pi\om}
\end{equation}
                if we introduce lattice spacing
        $ \Delta $
                to anticipate the continuous limit.

                Let us concentrate on the spin of
                the excitations. The results of section
        \ref{seight}
                are rather complicated. However they
                drastically simplify in the limit
        $s\to\infty$.
		Indeed the restriction
        $a\leqslant 2s$
		for kink labels disappears
                and the sequence
	$0, a_1,\ldots, a_{2n-1}, 0$
                can be considered as
                parametrizing a particular singlet,
                entering the representation of
        $\sltwo$
		in
	$\prod^{2n}\otimes\CC^2$.
		For
	$n=1$
                (two particle state) we have just one such state
	$(0,1,0)$,
		corresponding to a
                singlet in
	$\CC^2\otimes\CC^2$.
		The phase factor, corresponding to the triplet
                state in spin space and singlet in the
                kink space is given by the
        $s\to\infty$
		limit of
        (\ref{Slamb})
                and
        (\ref{Szlam})
                and is rather simple:
\begin{equation}
        S(\lam)=S_{\rm t}(\lam)
                \frac{\lam-i}{\lam+i}S_{\rm t}(\lam) \;\; ,
\end{equation}
                where
        $S_{\rm t}(\lam)$ is given by
                (\ref{Stlam}).
                The last expression is nothing, but the phase factor for a
	(triplet) $\otimes$ (singlet)
		state with respect to two independent
        $\sltwo$
		groups
                with
	$S$-matrix,
		given by a tensor product of two
	$S$-matrices of type
        $ S^{1/2,1/2} $.

                Now we mention, that
\begin{equation}
                \sltwo\otimes \sltwo = \, \ofour
\end{equation}
                with
	$\CC^2\otimes\CC^2$
		being a vector representation for it. In other
                words, we can interpret the excitations in
                our inhomogeneous
        XXX${}_s$
		model
                in the limit
        $s\to\infty$
		as vector particles corresponding to the vector
                representation of
        $\ofour$.

                This particle content coincides exactly
                with what is believed to be true in
                the
        $ \Spher^{3} $
		nonlinear
	$\si$-model
		(or the
        $ \sltwo $
                principal chiral model). The inhomogeneous
        XXX${}_s$
		spin model
                just realizes one particular sector of
        $\sltwo$
		chiral model
                in the limit
        $s\to\infty$;
		only particles in singlet state with respect to
                the right spin appear. However this is enough
                to characterize the full
	$S$-matrix.

                There are more reasons to justify this
                correspondence, in particular the
                exact calculation of the
        $\bet$-function
		of the renormalization group is
                possible via the BAE.
                I do not have time to discuss it here.

                The second example is the Sine-Gordon model.
                This dynamical system was
                instrumental for the developing the ABA
                by Sklyanin, Takhtajan and me in
                the end of 70-ties. From that time there were
                developed quite a few
                approaches to investigate it via BAE. Here
                I present an approach based on
                alternating inhomogeneous chain, developed by A.~Volkov
                and me at 1992.

                As the dynamical variables on an alternating lattice
                I shall use the Weyl
                variables
	$u_n$,
	$v_n$
		with the exchange relations
\begin{equation}
                u_n v_n=q v_n u_n,\quad q=e^{i\ga} \;\; .
\label{exchange}
\end{equation}

                The auxiliary Lax operator
\be
        L_{n,a}(x)=\left(
                       \begin{array}{cc}
                              u_n       &   x v_n\\
                             -xv_n^{-1}&u^{-1}_n
                       \end{array}
                    \right)
\label{Luva}
\ee
                belongs to the
        XXZ
                family with
        FCR
        (\ref{FCRZ}).
                I cannot help mentioning,
                that in simplicity formula
        (\ref{Luva})
                beats even the expression
                for the Lax operator of
        XXX
                model
        (\ref{Laxlam}).

                The local product of two Lax operators
        $L_{2n,a}(x\kappa)L_{2n-1,a}(x\kappa^{-1})$
                corresponds to transport along a
                physical lattice site.
                We have explicitly the matrix
\be
\left(
\begin{array}{cc}
        u_{2n}u_{2n-1}-x^2 v_{2n}v^{-1}_{2n-1} &
          x\left(
                \kappa^{-1} u_{2n}v_{2n-1}+\kappa v_{2n}u^{-1}_{2n-1}
          \right)\\
       -x\left(
             \ka v^{-1}_{2n}u_{2n-1}+\ka^{-1} u^{-1}_{2n}v^{-1}_{2n-1}
        \right) &
                u^{-1}_{2n}u^{-1}_{2n-1}-x^2v^{-1}_{2n}v_{2n-1}
\end{array}
\right) \; .
\label{twoLax}
\ee
                We can identify this matrix with the
        XXZ
                Lax operator of type
                (\ref{Laxchi})
                with the spin variables
        $q^{S^{3}}$, $S^{\pm}$
                given by
                (\ref{qSpins}), (\ref{qSpin})
                if we impose the constraint
\be
                \tilde w_n=u_{2n}u_{2n-1}v_{2n}v^{-1}_{2n-1}=1
\ee
                at each physical lattice site.
                This constraint reduces the number
                of local degrees of freedom from two to one,
                as it must be.

                Indeed, let us put
\ba
                e^{i\pi_{n}} & = & u_{2n}v_{2n-1} \;\; ;\\
                e^{i\vph_{n}} & = & u_{2n}u_{2n-1} \;\; .
\ea
                Then
        (\ref{twoLax})
                turns to
        (\ref{Laxchi})
                after division by
        $x$
                and similarity transform with matrix
\be
                D=\left(\begin{array}{cc} \kappa^{1/2}&0\\
                0&\kappa^{-1/2}\end{array}\right) \;\; ,
\ee
                if we put
\be
                m^2=\kappa^2 \; .
\ee

                After this identification
                it is natural to assume that the
        BAE
                look like
        (\ref{BAEch}).
                The derivation is nontrivial,
                because the representation for the
        $q$-deformed
                spin variables is not of the highest weight.
                However it can be done and the local factors
        $\al(\lam)$
                and
        $\dl(\lam)$
                correspond to spin
        $-1/2$.
                Returning to the additive variables
        $x=e^\lam$, $\kappa=e^{\om}$
                we have the
        BAE
                in the form
\be
        \left( \frac
        {       \sinh\left(
                        \lam_j+\om+\frac{i\ga}{2}
                   \right)
                \sinh\left(
                        \lam_j-\om+\frac{i\ga}{2}
                   \right)}
        {       \sinh\left(
                        \lam_j+\om-\frac{i\ga}{2}
                   \right)
                \sinh\left(
                        \lam_j-\om-\frac{i\ga}{2}
                   \right)} \right)^{N/2} =
                \prod^l_{{\scriptstyle
                                k=1 \atop k\ne j }}
         \frac
        {\sinh(\lam_j-\lam_k-i\ga)}
        {\sinh(\lam_j-\lam_k+i\ga)}
        \; .
\label{BAEdiscr}
\ee

                Now I turn to the description
                of the time shift operator.
                For that we are to find the fundamental Lax
        $L_{n,f}(x)$
                operator,
                corresponding to the auxiliary
        $L_{n,a}(x)$
                from
        (\ref{Luva}).
                We write the equation
\be
        L_2\left(1/x\right) L_1\left(1/y\right)
                r\left(x/y\right) =
                        r\left(x/y\right)
        L_1\left(1/y\right) L_2\left(1/x\right)
\label{LLr}
\ee
                with the natural brief notations
\be
                L_1 {(x)}=L_{n_1,a} {(x)},\quad
                        L_2(x)=L_{n_2,a}(x)
        \; .
\ee
                The solution will be looked for
                as a function of the variable
\be
                w = u_2 u_1 v^{-1}_2 v_{1}
        \; .
\ee
                The diagonal matrix elements of equations
        (\ref{LLr})
                are trivially satisfied.
                The right upper matrix element of the
                equation
        (\ref{LLr})
                looks as follows
\be
        (x u_{2} v_{1} + v_{2} u_{1}^{-1}) r(x,w) =
             r(x,w) (u_{2} v_{1} + x v_{2} u_{1}^{-1})
\ee
                and can be reduced to the functional equation
\be
        \frac{r(x,qw)}{r(x,q^{-1}w)}=\frac{xw+1}{x+w}
\label{rfunceq}
\ee
                by using the exchange relations
        (\ref{exchange}).
                We have already seen this equation
                in section
        \ref{sten}.
                Indeed, for the representation
        (\ref{qSpins}), (\ref{qSpin})
                we can check, that
        $w$
                coincides with
        $q^{2J+1}$.

                Denoting
\ba
                w_{2n}&=&u_{2n}u_{2n-1}v^{-1}_{2n}v_{2n-1}
                \quad ; \\
                w_{2n+1}&=&u_{2n+1}u_{2n}v^{-1}_{2n+1}v_{2n}
        \; ,
\ea
                we have the expression
                for the time evolution operator
\be
                e^{-iH}=\prod r(\ka^{2},w_{\even})
                        \prod r(\ka^{2},w_{\odd })
        \; .
\label{eiH}
\ee

                This expression is simple enough
                to allow us to write down
                the equation of motion explicitly.
                It is convenient to introduce a
                new set of variables
        $\psi_n$,
                such that
        $w_n$
                are their
                ``multiplicative derivatives''
\be
                w_n=\frac{\psi_{n+1}}{\psi_{n-1}}
        \; .
\label{wpsi}
\ee
                Variables
        $\psi_{m}$
                and
        $\psi_n$
                commute for
        $n-m$ even, so
                there is no problem in order of factors in
                (\ref{wpsi}).
                The exchange relations among
        $\psi_n$
                are nonlocal, but it is not of concern to us;
                it suffices to say, that given
        $\psi_n$
                does not commute with only one
        $w$,
                namely
        $w_n$,
                and for this pair there is the Weyl relation
\be
                \psi_n w_{n} = q^2 w_n \psi_n
        \; .
\ee
                We can take the set
        $\{\psi_{\even}, w_{\even}\}$
                or
        $\{\psi_{\odd}, w_{\odd}\}$
                as a set of independent Weyl
                variables in the physical Hilbert space.
                I skip here the subtle question
                of boundary conditions for integrating
                (\ref{wpsi}),
                which is settled in the original literature.

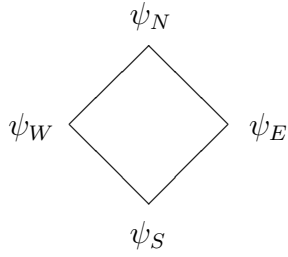
\begin{figure}
\begin{center}
\begin{picture}(120,110)(0,0)
  \multiput(30,60)(30,-30){2}{\line(1,1){30}}
  \multiput(30,60)(30,30){2}{\line(1,-1){30}}
  \put(7,56){$\psi_W$}   \put(53,98){$\psi_N$}
  \put(53,15){$\psi_S$}   \put(98,56){$\psi_E$}
\end{picture}
\end{center}
\caption{Elementary plaquette}
\label{elplaq}
\end{figure}

                Equations of motion are especially simple in
        $\psi$
                variables.
                Let
        $\hat\psi_n$
                be a variable
        $\psi_n$
                once displaced in time
\be
                \hat\psi_n=e^{iH} \psi_n e^{-iH} \quad .
\ee
                Take for definiteness
        $\psi_{\odd}$,
                i.e.
        $\psi_{2n+1}$.
                Only one factor, containing
        $w_{2n+1}$
                in
        (\ref{eiH}),
                does not commute with it,
                so we have using the functional equation
        (\ref{rfunceq})
\ba
        \hat\psi_{2n+1} & = &
           r^{-1} (w_{2n+1}) \,
                \psi_{2n+1} \, r(w_{2n+1})
                                =
                \psi_{2n+1}r^{-1}(q^{-2}w_{2n+1})r(w_{2n+1}) =
\nonumber \\
                               & = &
         \psi_{2n+1} \frac{\ka^2q^{-1}w_{2n+1}+1}
                          {\ka^2 + q^{-1}w_{2n+1}} =
         \psi_{2n+1} \frac{\ka^2q^{-1}\psi_{2n+2}+\psi_{2n}}
                          {q^{-1}\psi_{2n+2}+\ka^2\psi_{2n}}
        \; .
\label{eqmot}
\ea
                This is our equation of motion. For
        $\psi_{\even}$
                the derivation is exactly the same.
                On the discrete space--time drawn
                on Figure
        \ref{insaw} the equation
        (\ref{eqmot})
                connects the
        $\psi$
                attached to an elementary plaquette
                shown in Figure
        \ref{elplaq}
                \noindent  and can be rewritten as
\be
                \psi_N=\psi_S
          \frac{\ka^2q^{-1}\psi_{W}+\psi_{E}}
                  {q^{-1}\psi_{W}+\ka^2\psi_{E}}
\label{psieqmot}
\ee
                or
\be
        (q^{-1}\psi_{N}\psi_{W} - \psi_{S}\psi_{E} ) =
        \ka^2 (q^{-1}\psi_{S}\psi_{W} - \psi_{N}\psi_{E} )
        \; .
\label{sgeqmot}
\ee

                The last simple quadratic equation
                looks quite appealing.

                Let us show, how the Sine--Gordon equation
                appears in the formal continuous limit.
                We also confine ourselves to classical
                commuting variables.

                The naive prescription that
        $\psi_n$
                define a continuous function
                of space--time variables does not work.
                We have already seen a similar situation
                in the case of
        NLS
                model.
                The way out is to use variables
\be
                \chi=
                        \left \{
                        \begin{array}{c}
                                \psi\\
                                \psi^{-1}
                        \end{array}
                        \right.
\ee
                alternatively on each second
        $SE$
                characteristic line of our
                lattice. Then equations of motion
        (\ref{psieqmot})
                transform into
\be
                \chi_{N}= \chi_{S}^{-1}
          \frac{\ka^2q^{-1}\chi_{W}\chi_{E}+1}
                  {q^{-1}\chi_{W}\chi_{E}+\ka^2}
\ee
                and for large
        $\ka^2$
                and classical limit
        $ q = 1 $
                we have
\be
        \frac
        {\chi_{N}\chi_{S}}
        {\chi_{E}\chi_{W}}
                =1+
                \frac1{\ka^2}
        \left(\frac1{\chi_{E}\chi_{W}}-\chi_{E}\chi_{W}
        \right)+\ldots
        \; .
\label{chiNS}
\ee
                We put here
\be
                \chi=e^{i\vph}
\ee
                and consider
        $\vph$
                as defining a smooth function of
        $x,t$
                in the continuous limit. Then evidently
\be
        \frac{\chi_{N}\chi_{S}}
             {\chi_{E}\chi_{W}}
                =\exp\left\{i\frac{\Dl^2}{2}
                        (\vph_{tt}-\vph_{xx})+\ldots
                     \right\}
\ee
                and introducing the rescaling
\be
                \frac1{\ka^2}= m^2\Dl^2
\label{scale}
\ee
                we get from
        (\ref{chiNS})
                the Sine--Gordon equation
\be
                \vph_{tt}-\vph_{xx}+2m^2 \sin 2\vph=0
        \; .
\ee
                In quantum version the scaling
        (\ref{scale})
                is to be modified,
                accounting to the mass renormalization.
                The investigation of
        BAE
        (\ref{BAEdiscr})
                allows to describe this renormalization
                exactly, adding to the nonperturbative
                intuition in Quantum Field Theory.
                However I cannot speak about it here
                and stop technical developments to come
                to some conclusions.

\section{ Conclusions and perspectives }
\label{sthirteen}

		The first and the last formulas in this
		text being the same, our exposition
		closed the circle. We have learned the
		technique of Algebraic Bethe Ansatz
		for solving integrable models and have shown,
		how it works in detail on the
		simplest example of spin 1/2 XXX magnetic chain.
		Several other models were
		treated more superficially, only the
		specific details were given. Several
                parameters, appearing in these
		generalizations: spin
	$s$,
		anisotropy parameter
	$\ga$,
		shift
	$\om$
                in the alternating chain, allow to include
		in our treatment most known examples
		of soliton theory, including
		relativistic model of Quantum Field Theory.
		Thus the spin chains showed
		their great universality. If we add,
		that the fashionable models of
		conformal field theory also can be
		included as particular limits, then we
		could claim, that we deal with the
		classifying object in the theory of
		integrable models.

		We treated here only rank 1 case associated
		with the Lie algebra
        $\sltwo$.
		Generalization to the higher rank simple
		Lie algebras is
		possible and many results are already known.
		In particular, one can use a
		Cartan subalgebra label to introduce
		one extra dimension for space.

		We did not make a stress on the subject
		of quantum groups or quantum
                symmetries, as this was discussed in other
                lectures at the School.
		However I suppose, that I made it clear,
		how quantum groups were created
		naturally inside of ABA (of course,
		their appearance in the $\CC^*$ algebra
		approach of Professor Woronowitz
		was completely independent).

		Discrete space-time approach to quantum
		integrable models seems to me most
		promising and elegant. It has already allowed
		to write explicitely the quantum
                equation of motion. I cannot help feeling
		that it is just the beginning of
		analysis, including complex
		analysis, on discrete manifolds.

		From pure mathematical point of view
		many of reasonings in my lectures were
		only heuristic. The statements were ``claims''
		rather than ``propositions''
		not to mention ``theorems''. I am not
		a supporter of absolute rigor in
		mathematical physics, but would
		be happy, if some of my claims become
		better justified. Even more important is
		the fact, that new mathematical
		constructions and results could appear
		in course of this justification.
		Especially promising are problems of
		infinite tensor products in
		antiferromagnetic case and their connections
		with infinite--dimensional Lie
		algebras and analysis on the discrete
		space-time. I leave these challenges to
                my listeners.

\section{ Comments on the literature on BAE }
\label{sfourteen}

		Here I will add some comments which will
		serve as a guide to the literature
		on Bethe Ansatz. Only  seminal
		(from my own point of view) papers will be
		mentioned. The recent monograph
        \cite{QScat}
		contains quite vast list of
		references. Another monograph
        \cite{Gandin}
		reflects the developments around Bethe
		Ansatz mostly prior to the advent of
		ABA.

		BAE were first written by H.~Bethe in
        \cite{Physik}
		for the spin 1/2 XXX model. He
                did not use the uniformizing rapidity
		variables, which apparently appeared
		first in a paper by Takahashi
        \cite{Takah}.
		First integral equation for the
		distribution of roots was derived by
                L.~Hulthen
        \cite{Hult}.
		My exposition of the
		spin 1/2 XXX model follows closely
		my paper with Takhtajan
        \cite{TakhFad81}.
		The
		statement, that the spin of a spin wave
		is 1/2 was announced by us in
        \cite{Takht}.
		The scattering theory for the excitations
		over the ferromagnetic vacuum was
		developed by Babbit and Thomas
        \cite{Babbitt}.

		General formalism of the ABA was worked out
		by Sklyanin, Takhtajan and me in
        \cite{Sklyan79}
		on the example of the Sine--Gordon model.
		The state of art at the
		end of 70--ties was described in my survey
        \cite{Fad80}.

		The $q$--deformed
        $\sltwo$
		algebra with defining relations
        (\ref{Sq}), (\ref{Spm})
		was introduced by Kulish and Reshetikhin
		in 1981 in
        \cite{KulishResh81}
		in connection with
		the higher spin XXZ model. More general
		investigation of the quadratic algebras
		was performed by Sklyanin
        \cite{Sklyan83},
		who pointed out the connection with the Hopf
		algebras
        \cite{Sklyan85}.
		The proper algebraic understanding of
		general structures,
		appearing here, was formulated by Jimbo
        \cite{Jimbo85}
		and in the most perfect form by Drinfeld
        \cite{Drin87}.
		The role of the fundamental Lax operator
		was underlined by Tarasov,
		Takhtajan and me in
        \cite{TakhFad83}.
		There we followed the ideas of fusion,
		introduced by
		Kulish, Reshetikhin and Sklyanin in
        \cite{KulishReshSklyan81},
		where the formula
        (\ref{rJlam})
		was derived.

		Investigation of BAE for higher spin
		XXX model was done by Takhtajan
        \cite{Takht82},
		(see also Babujan
        \cite{Babujan83}).
		The kink interpretation of the
		excitations was
		developed by Reshetikhin in
        \cite{Resh91}.
		The thorough investigation of BAE for NLS
		model with chemical potential was
		done by Lieb and Liniger
        \cite{Lieb63}.

		The interpretation of the NLS and
		SG models as XXX and XXZ chains,
		correspondingly, was done by Izergin
		and Korepin
        \cite{Izerg84}.
		Korepin was also the
		first to propose the formula of type
        (\ref{Stlam})
		for the matrix element of
	$S$-matrix
		for the excitations above Dirac sea
        \cite{Korep??}.

		The functional equation for the
		fundamental Lax operator
        (\ref{funleq})
		for XXZ model has a natural place
		in the theory of
        $\slaff$
		algebra, see e.g.
        \cite{Frenkel92}.

		The idea of imbedding the
        $\Spher^2$ $\si$-model
		into XXX chain belongs to
		Haldane
        \cite{Haldane83}
		and Takhtajan and me
        \cite{TakhFad83pp}.
		More detailed development was done
		by Affleck, who in particularly stressed
		the role of the topological
        $\vartheta$-term
        \cite{Affleck89}.
		The BAE
        (\ref{BAEcorr})
		was derived by Bytsko (unpublished).

		The idea of the usefulness alternating
		inhomogenous chain to treat the
		relativistic models was
                underlined by Reshetikhin and me
        \cite{FadResh86}
		in our
		treatment of the principal
        $\sltwo$
		chiral model. It was
		developed further by Destri and De-Vega
        \cite{Destri}.
		The zero curvature
		interpretation was given by Volkov and me in
        \cite{VolkFad92}.

		The quantum equations of motion
        (\ref{sgeqmot})
		for the SG model appeared in
        \cite{FadVolk94};
                they coincide in the classical limit
	$(q=1)$
		with Hirota equations
        \cite{Hirota77}.
		In terms of variables
	$f_A$,
		connected with
	$\psi_A$ by
\be
	f_A-f_B=\frac1{\psi_A\psi_B}
\ee
		in the form
\be
	\frac{(f_N-f_E)(f_W-f_S)}
                {(f_N-f_W)(f_E-f_S)}=\kappa^4 \quad ,
\ee
		as was commented by Volkov
        \cite{Volkpp}.
		This makes the appearance of discrete
		complex analysis indispensable. The
		same equations were recently
                advertised by Capel et al.
        \cite{Capel95}.
		Another line of
		thought on discrete geometry and
		discrete SG equation belongs to Pinkal
                et al.
        \cite{Bobenko93}.
                Quadratic algebras appear now in many
		instances and guises. I
		mention the discretized affine algebra,
		introduced by Semenov-Tjan-Shansky
		and discussed in some detail in
        \cite{Fad92}.

		I finish by reference to my previous
		lecture course
        \cite{Fad95}
		where many
		aspects of these lectures were treated,
                and to lecture notes of my
		Schladming lectures
		on new applications of BAE to
		Hofstadter model
                from condensed matter physics,
		to high energy QCD and to Liouville
		model of Conformal Fiel Theory.

		In my lectures I did not refer at all
		to the parallel development in
		classical statistic physics.
		The general equivalence exists between 1+1
		dimensional quantum dynamical systems
		and 2 dimensional models of classical
		statistical physics. Integrable models
		of the former domain correspond to
		the exactly soluble models in the latter.
		A lot of well known results here are
                connected with the names of Onsager,
		C.~N.~Yang, Lieb, Baxter and others. I
		can refer to the monograph of Baxter
        \cite{Baxter82},
		where this theme is displaced in
		great detail.

		Another connection with statistical
		physics is the use of the integrable
		hamiltonians to define the
		corresponding Hibbs state
        $ Z^{-1} e^{-\bet H+\mu N}$.
		This was pioneered in paper of
		C.~P.~Yang and C.~N.~Yang
        \cite{YYang69},
		which led to important development,
		called the Thermodynamic Bethe Ansatz.
		Already mentioned paper
        \cite{Babujan83}
		of Babujan was an important step in
		formulating this technical development.

		Returning to quantum field theory
		interpretation I must say, that until now
		ABA has given in the way to investigate
		mass spectrum and
	$S$-matrix. More
		detailed off shell characteristics
		of systems under consideration are very
		scarce. The main success was achieved
		by Smirnov in the discussion of
		formfactors of local operators
        \cite{Smirnov92}.

		Another line of thought, originated
		by Korepin, is partly described in
        \cite{QScat}.
		This topic is the most important from
		the point of view of mathematical
		physics and physical applications.

\end{document}